\documentclass[11pt,reqno]{amsart}

\usepackage{amsmath, amsthm, amssymb,amsaddr}
\usepackage{mathtools}
\usepackage{algpseudocode}
\usepackage{algorithm}
\usepackage{subfigure}
\usepackage{epstopdf}
\usepackage{tikz}
\usepackage{color}
\usepackage[export]{adjustbox}
\usepackage{caption}
\usepackage{hyperref}

\theoremstyle{remark}
\newtheorem{remark}{Remark}

\newtheorem{lemma}{Lemma}
\newtheorem{definition}{Definition}

\DeclareMathOperator*{\argmin}{arg\,min}

\newcommand{\be}[1]{\begin{equation}\label{#1}}
\newcommand{\ee}{\end{equation}}

\newcommand{\no}[1]{#1}

\newcommand{\bx}{\boldsymbol{x}}
\newcommand{\bd}{\boldsymbol{d}}

\renewcommand{\d}{\mathrm{d}}

\renewcommand{\no}[1]{} 

\title[Three-dimensional ray-based ultrasound tomography]{Refraction-corrected ray-based inversion for three-dimensional ultrasound tomography of the breast}     
\author{A.\ Javaherian, F.\ Lucka$^*$ \& B.T.\ Cox}
\email{a.javaherian@ucl.ac.uk}
\address{Department of Medical Physics \& Biomedical Engineering,\\ University College London, London, UK. WC1E 6BT\\
$^*$Centrum Wiskunde \& Informatica (CWI),\\ 1098 XG Amsterdam, The Netherlands, \\  
$^*$Centre for Medical Image Computing,\\ University College London, London, UK. WC1E 6BT}

\date{\today}

\begin{document}
\maketitle

\begin{abstract}
Ultrasound Tomography has seen a revival of interest in the past decade, especially for breast imaging, due to improvements in both ultrasound and computing hardware. In particular, three-dimensional ultrasound tomography, a fully tomographic method in which the medium to be imaged is surrounded by ultrasound transducers, has become feasible. This has led to renewed attention on ultrasound tomography image reconstruction algorithms. In this paper, a comprehensive derivation and study of a robust framework for large-scale bent-ray ultrasound tomography in 3D for a hemispherical detector array is presented. 
Two ray-tracing approaches are derived and compared. More significantly, the problem of linking the rays between emitters and receivers, which is challenging in 3D due to the high number of degrees of freedom for the trajectory of rays, is analysed both as a minimisation and as a root-finding problem. The ray-linking problem is parameterised for a convex detection surface and three robust, accurate, and efficient ray-linking algorithms are formulated and demonstrated. To stabilise these methods, novel adaptive-smoothing approaches are proposed that control the conditioning of the update matrices to ensure accurate linking. The nonlinear UST problem of estimating the sound speed was recast as a series of linearised subproblems, each solved using the above algorithms and within a steepest descent scheme. The whole imaging algorithm was demonstrated to be robust and accurate on realistic data simulated using a full-wave acoustic model and an anatomical breast phantom, and incorporating the errors due to time-of-flight picking that would be present with measured data. This method can used to provide a low-artefact, quantitatively accurate, 3D sound speed maps. In addition to being useful in their own right, such 3D sound speed maps can be used to initialise full-wave inversion methods, or as an input to photoacoustic tomography reconstructions.
\end{abstract}

\clearpage
\tableofcontents
\clearpage

\section{Introduction}
\label{sec:Introduction}
Ultrasound Tomography (UST) has received growing interest in the past decade, especially for breast imaging \cite{Li1,Wiskin1,Wiskin2,Ruiter2,Hopp1,Op}. Despite first being proposed nearly 50 years ago \cite{Green}, it was not until recently that improvements in both ultrasound (US) and computing hardware have begun to be exploited to bring UST to the point where it can potentially compete with, and complement, the more established medical imaging modalities. UST is a fully tomographic method in which the medium to be imaged is surrounded by transducers (emitters and receivers) and US is sequentially transmitted from the transmitters (or from small groups) into the imaging target. 
The US propagates through the tissue, is scattered, refracted and attenuated, and finally detected by the array of detectors. The inverse problem is to recover images of the acoustic properties of the target. 
These typically include images of sound speed or acoustic attenuation, which can be quantitative, and reflection images (which are usually qualitative but related somewhat to the gradient of the acoustic impedance). 
The hope is that this quantitative information about the tissue properties can be used to aid diagnosis \cite{Ruiter}. (It should be noted that UST is very different from conventional ultrasound imaging as widely-used in clinical settings, which is a backward-mode imaging modality that uses a small-area probe to give qualitative reflection images in real time.) The majority of UST approaches have been limited to 2D cases, and thus out-of-plane effects are neglected, leading to image artefacts. This study concentrates on 3D volumetric reconstruction of the sound speed distribution from measurements made on a hemispherical surface around the imaging target.

The UST inverse problem can be tackled using full-waveform inversion approaches, such as are used in seismic applications, which are based on a minimisation of the $L_2$-norm of the discrepancy between the measured acoustic pressure and that from a numerical model \cite{Wang,Matthews,Matthews1,Bachmann2020,Liva}. To do this, the acoustic pressure is modelled using the wave equation for heterogeneous media, and the unknown acoustic property (here the sound speed) is iteratively updated using a computation of gradient of the $L_2$-norm of the discrepancy \cite{Plessix}. This class of methods can provide high spatial resolution images, are flexible, in the sense that different forward models can be used, and make use of all the information in the measured data. 
(See, for example, \cite{Wang,Matthews,Matthews1} for time-domain methods and \cite{Wiskin1,Wiskin2} for frequency-domain approaches.) 
However, one problem with these approaches is that they are very computationally expensive, and tackling 3D imaging scenarios remains challenging \cite{Wang,Matthews,Matthews1}. They are also vulnerable to uncertainties in the forward modelling, e.g., in the transducer properties \cite{Huth}, and the solution tends to converge to the nearest local minimum instead of the global minimum of the problem.
As a result, minimisation-based full-wave approaches typically require a good initial guess. There is therefore still a great deal of interest in other approaches to UST, especially methods that are less intensive computationally.

Ray-based methods, which use a high-frequency approximation to reduce the wave propagation problem to a ray-propagation problem, are still popular. Indeed, ray-based methods are commonly used to provide the initial guess for full-wave inversions \cite{Wiskin1,Wang,Matthews1,Liva,Wiskin2}. This approach, initially inspired by X-ray tomography, uses measurements of the time-of-flight (TOF) of the acoustic signals across the imaging target to reconstruct the slowness distribution (the reciprocal of the sound speed) using Radon-type inversion techniques \cite{Anderson1}. 
Typically, two sets of data are recorded, one from a known homogeneous reference medium such as water and another from the target. The inverse problem then becomes reconstructing the difference in the slowness distributions between the target and water from the differences in the TOFs measured between every pair of emitter and receiver \cite{Duric}. 
To solve this inverse problem, a method of relating the TOFs to the slowness is required. This is usually achieved by computing the TOFs as the integrals of the slowness along the rays that link the pairs of emitters and receivers. The inverse problem now becomes a minimisation of the norm of the discrepancy between the measured and modelled TOFs via an iterative adjustment of the slowness distribution. This is often done using a successive enforcement of a Radon-type forward operator and its adjoint using the trajectories of the rays \cite{Li1}. This raises the question of how the ray trajectories are computed. 

Early approaches to UST used an assumption that the acoustic waves propagate along straight lines \cite{Duric}, thus neglecting diffraction, refraction and scattering \cite{Denis}. Refraction (bending) of the rays can be incorporated by basing the calculation of each ray trajectory on Fermat's principle, which states that the energy tends to propagate along the path with minimal acoustic length \cite{Anderson2,Ferra, Denis,Oliveira}. However, because rays inherently assume a high-frequency approximation, diffraction and scattering effects are still neglected \cite{Denis}.

One class of ray-tracing approaches is based on tracking the wavefront from an emitter across the medium to all receivers simultaneously \cite{Klimes}. Among them, the fast marching method (FMM) \cite{Tsitsiklis,Sethian} has received much attention for UST \cite{Li2,Ali}. However, because this approach is based on a calculation of the wavefronts through the entire domain, it becomes computationally intensive for the 3D inverse problem, and has predominantly been used for 2D studies. For example, in \cite{Ali}, FMM was used to compute TOFs for a 2D medium encompassed by a circular detection surface. In \cite{Li2}, out-of-plane refraction is accounted for by applying the FMM to a slab-like volume including the target slice aligned with the circular transducer array.

Another class of ray-tracing approaches, \textit{two-point ray tracing}, treats the emitters and receivers as ray-emission and ray-reception points and computes only the rays between them (for all emitter-receiver pairs). The numerical methods for calculating the trajectory of a ray that links a pair of emission and reception points can be categorised into \textit{bending}  \cite{Wesson,Julian,Pereyera,Norton1,Moser,Xu} and \textit{shooting} \cite{Anderson2,Anderson3,Oliveira} approaches.
In the bending method, the problem is framed as a two-point boundary value problem. By fixing both end-points of the ray, its trajectory is iteratively computed until the perturbation in the path becomes smaller than a tolerance \cite{Wesson,Julian,Pereyera,Norton1,Um,Moser,Xu}. In \cite{Wesson} and \cite {Julian}, central finite difference approaches were used for solving a second-order differential equation derived from an Euler Lagrange equation describing the trajectory of the rays
(see Sec.\ \ref{subsec:Ray_equation}).

In \cite{Pereyera}, a method for ray bending based on a reduction of the second-order ray equation to a set of first-order equations was proposed, and was numerically solved using adaptively-varying finite difference methods and varying meshes. It was shown that bending methods are more efficient than shooting methods for a simple distribution of the sound speed, but they are less efficient, or may fail for more complex media \cite{Cerveny}. In the shooting method, the ray's path is solved as a sequence of initial value problems, given the starting position and an initial direction for the ray. With the initial position of the ray fixed at the emission point, the initial direction of the ray is iteratively controlled until the end point of the ray intercepts the reception point to within a small tolerance. This is known as the \textit{ray-linking} problem \cite{Anderson2,Rawlinson1,Rawlinson2}. Shooting methods have been widely used for 2D UST imaging \cite{Anderson2,Anderson3,Oliveira}, but little work has been done on their extension to 3D.

Both classes of two-point ray-tracing approaches, \textit{shooting} and \textit{bending} methods, are non-linear inverse problems, and are subject to instability or divergence, especially for 3D scenarios \cite{Thurber,Pereyera}. For example, when the sound speed is non-differentiable or causes shadow zones \cite{Pereyera,Rawlinson2}, the data may no longer depend continuously on the unknown function, leading to ill-posedness in the Hadamard sense. This is easier to deal with in shooting methods, as testing the convergence of (and setting a convergence criterion for) a shooting method is straightforward, because the error functional is a measure of the distance between the end point of the ray and the reception point.

One further comparison between full-wave and ray-based approaches relates to rotating measurement systems. The computational load of full-wave inversions, or ray-based approaches based on wavefront tracking, does not change as the number of receivers increases, as it only scales with the number of emitters. While this can be an advantage, it is not necessarily so. For example, consider a fixed array with $N$ transducers, which act both as emitters and receivers. The number of unique emitter-receiver pairs will be $N(N-1)/2$. For some applications, including breast imaging, it may not be possible to fit enough transducers in the array because of their finite size, and it is necessary to rotate or translate the system to obtain sufficient data \cite{Ruiter2}. In this case, the number of effective emitters is $N_{\text{eff}} = N_{\text{rot}} N'$, where $N_{\text{rot}}$ is the number of rotations, and $N'<N$ is the number of transducers. The number of unique emitter-receiver pairs is given in the rotating case by $N_{\text{eff}}(N'-1)/2$, so for the rotating system to measure as much data as the equivalent fixed system, it is necessary for $N_{\text{eff}} > N$. In other words, data collected using a rotating array, as opposed to an equivalent fixed array, leads to an increased computational burden during the reconstruction when using full-wave or wavefront methods. For two-point ray tracing approaches, however, it will remain the same.

In this paper, a two-point ray-tracing approach for 3D transmission UST of the breast using a hemispherical detection surface will be described. Key to this is a novel technique to solve the problem of 3D ray-linking in this geometry. To the best of our knowledge, this manuscript presents the first refraction-corrected ray-based inversion approach for full-3D transmission UST of the breast. Although, here, the ray-linking approach is applied to a hemispherical detection surface, it is parameterised in a general fashion and can be straightforwardly used for any arbitrary convex detection surface. In Section \ref{sec:Inverse_Problem}, the inverse problem of UST is defined, based on a refraction-corrected ray-based approach for a hemi-spherical detection surface. In Section \ref{sec:Ray_tracing}, the ray tracing approach used is introduced, and novel efficient approaches to solving the associated inverse problem of ray-linking are derived in Section \ref{sec:Two-point_ray_tracing_ray_linking}. In Section \ref{sec:Numerical_results}, all aspects of the method - ray-tracing, ray-linking, and quantitative sound speed imaging - are numerically validated. Section \ref{sec:Discussion_Conclusions} contains a discussion and the conclusions. In the Appendix, 
the approach used for calculation of the TOFs (a slight modification of \cite{Li3}) is presented. 

\section{Inverse Problem of Ultrasound Tomography}
\label{sec:Inverse_Problem}
In this section, the forward operator under consideration and the associated inverse problem, are introduced.
\subsection{Forward operator}     
\label{subsec:Forward_operator}
Let $\bx = \big(x^1,...,x^d\big)$ denote a spatial position in $\mathbb{R}^d$ with $d $ the dimension. For UST, $d$ can be either $2$ or $3$, but here is restricted to $d=3$. $\Omega \subset \mathbb{R}^3$ is an open bounded set in $\mathbb{R}^3$ given by the hemispherical volume,  centred at the origin, which is bounded by the surface $\mathbb{S} =  \big( \mathbb{S}_1 \cup \mathbb{S}_2 \big) \subset \mathbb{R}^2$, where $\mathbb{S}_1 = \left\{ x^3 <0  \  \cap \ \mid \bx \mid = R \right\}$ is hemispherical surface of radius $R$, and $\mathbb{S}_2 = \left\{ x^3 = 0  \ \cap \  \sqrt{x^{1^2}  + x^{2^2}} \leqslant  R \right\}$ is a circular plane. 
$\Omega$ contains the spatially-varying part of the refractive index distribution, $ n(\bx) = c_w/c(\bx) $, where $c(\bx)$ denotes the spatially varying sound speed distribution, and $c_w$ is a scalar value representing the reference sound speed (here the sound speed in water), i.e.\ $(n(\bx)-1)\in C_0^{\infty}(\Omega)$.

\begin{definition} 
\label{def:ray}
A ray is defined using $f_{ (n;e,p)} (\bx) = 0$, where $e \in \mathbb{S}_1$ denotes an initial point for the ray, and $p \in \mathbb{S}$ represents the interception point of the ray with surface $\mathbb{S}$ after traveling through the medium.
\end{definition}

\begin{figure}[htb]
    \centering
	\includegraphics[width=0.4\textwidth]{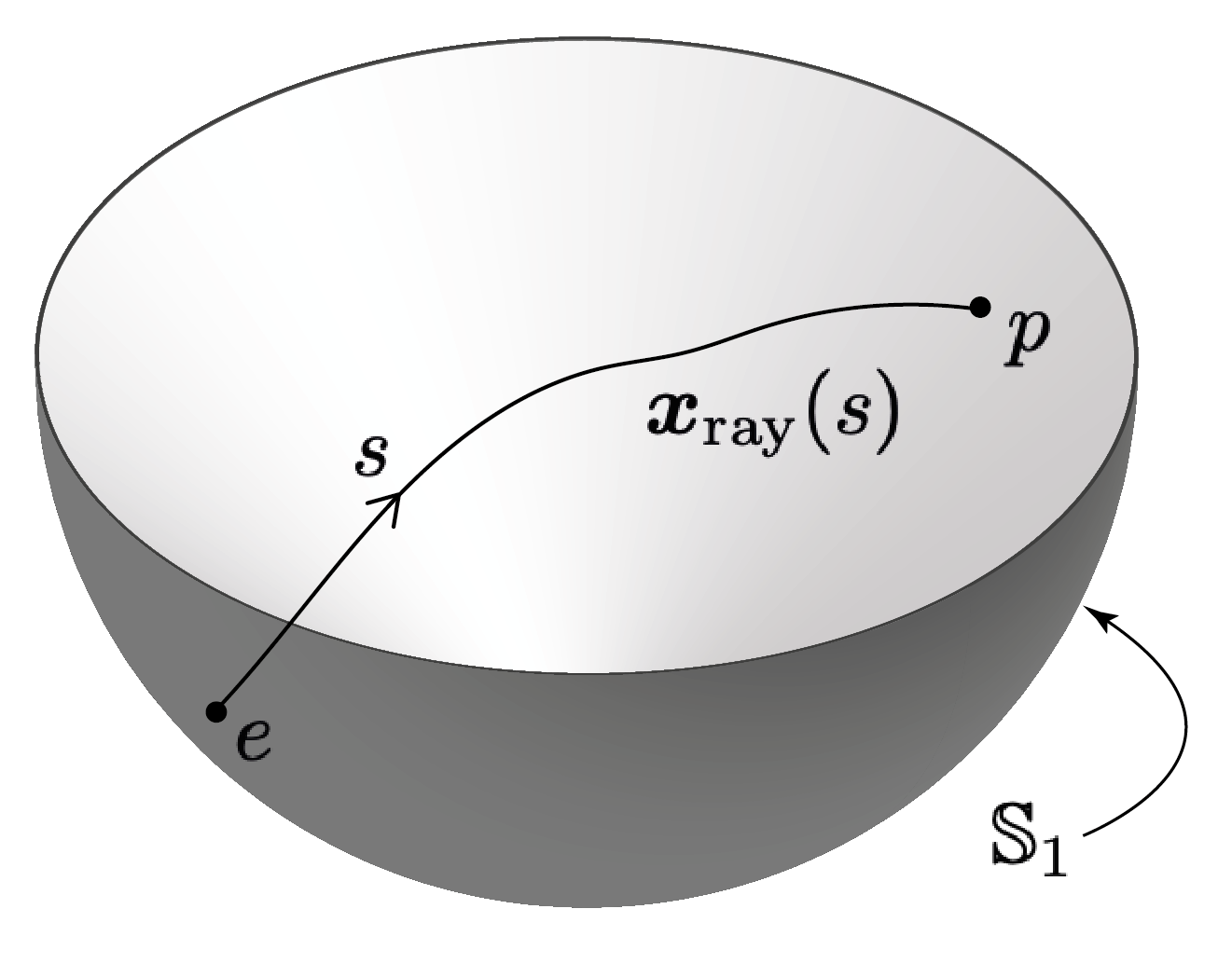}
    \caption{A ray starts at the emission point $e \in \mathbb{S}_1$ (also written as in $\bx_e \in \mathbb{R}^3$) and ends at the interception point $p$ (or $\bx_p$). The ray consists of the points ${\bx_{\text{ray}}(s),s\in [0, L_{\text{ray}}]}$ that satisfy $f_{(n,e,p)}=0$ (see Definitions 1 and 2).}
    \label{fig:ray_definition}
\end{figure}

\begin{definition}
The continuous forward operator is defined as 
\begin{align} 
\label{eq:forward_operator}
\begin{split}
& \mathcal{A}: \mathbb{D} (\Omega) \rightarrow \mathbb{S}_1 \times \mathbb{S} \\
& L_{(e,p)}= \mathcal{A}_{(n;e,p)} \left[ n(\bx) \right] = \int_{\Omega} n(\bx)   \delta\big(f_{(n;e,p)} (\bx) \big) \ d\bx,
\end{split} 
\end{align}
where the space $\mathbb{D}$ is defined such that any function $n(\bx) \in \mathbb{D}$ satisfies the condition that $(n(\bx)-1) \in C_0^{\infty}(\Omega)$.
Here, $\delta $ denotes the Dirac delta function, and $L_{(e,p)}$ is the acoustic length along a ray that links the point $e$ to the point $p$, and is given by the path integral of $n(\bx)$ along the ray $\delta\left(f_{(n;e,p)} (\bx)\right) $.
(A ray can be parameterised by the distance along it, $s$, by describing it as the line of points $\{ \bx_{\text{ray}}(s), s\in[0,L_{\text{ray}}]\}$ that satisfy $f_{(n,e,p)}(\bx)=0$, where $L_{\text{ray}}$ is the physical length - not the acoustic length - of the ray; see Fig.~\ref{fig:ray_definition}.)
\end{definition}

\subsection{Discrete forward operator}     
\label{subsec:Discrete_forward_operator}
In practice, the calculations are done on a grid of $N_n$ discrete grid points. $\bar{n} \in \mathbb{R}^{N_n}$ refers to the discretised refractive index field defined on the grid, $\bar{e} \in  \mathbb{S}_1 $ and $ \bar{r} \in \mathbb{S}_1 $ indicate the discretised emission and reception points, and $N_e$ and $N_r$ are the number of emitters and receivers, respectively. 
Note that the discretised emission and reception points do not necessarily lie on the grid points but rather correspond to the centres of actual transducer positions, as defined in the experiment. (In general, an overbar, $\bar{\cdot}$ indicates a discretised variable.) The measurements consist of the times-of-flight $\bar{T} \in \mathbb{R}^{N_e N_r}$ between every physical emitter and receiver.

\begin{definition} 
\label{def:discrete_forward_operator}
A discretised variant of the forward operator can be defined as
\begin{align}  
\label{eq:discrete_forward_operator}
\begin{split}
A & : \mathbb{R}^{N_n} \rightarrow \mathbb{R}^{N_e N_r} \\
\bar{L} & = A \big[ \bar{ n} \big]  ,
\end{split}
\end{align}
where $ \bar{L} = c_w \bar{T}$ is the discretised acoustic length, and, from Eq.\ \eqref{eq:forward_operator}, the dependence of $\bar{L}$ on $\bar{n}$ is nonlinear. Also, $\bar{L}_{(\bar{e},\bar{r})}$ is used to denote the acoustic length between an emission point $\bar{e}$ and reception point $\bar{r}$. 
(Note that a ray in the continuous domain links an arbitrary point $e \in \mathbb{S}_1$ to an interception point $p \in \mathbb{S}$, and a ray in the discretised domain links a discretised emission point $\bar{e} \in \mathbb{S}_1 $ to a discretised reception point $\bar{r} \in \mathbb{S}_1 \subset \mathbb{S}$.) The emitters and receivers are assumed to be points in space and emit and receive rays in all directions with equal weighting.
\end{definition}

\begin{definition}   
\label{def:Jacobian}
A linearisation of $A$ around a fixed refractive index distribution $\bar{n}$ gives a Jacobian matrix of the form 
\begin{align}
\label{eq:Jacobian}
\begin{split}
 &   J  \in  \mathbb{R}^{{N_e N_r}  \times N_n }   \\
 & \delta \bar{L} =   J \big[ \bar{ n} \big]  \delta \bar{n},
\end{split}
\end{align}
where $\delta \bar{L}$ denotes the change in the acoustic length due to perturbation $ \delta \bar{n} $ in the refractive index field.
The action of $J$ on $\delta \bar{n} $ for each pair of emitters and receivers is equivalent to an inner product of the corresponding row in $J$ by $\delta \bar{n}$, and gives the integral of the refractive index distribution along a ray that links this pair. Further details are given in Section \ref{subsec:Grid-to-ray_interpolation}.
\end{definition}

\subsection{Inverse problem}       
\label{subsec:Inverse_problem}
The UST inverse problem is to find the refractive index distribution of an object $\bar{n}_{\text{object}}(\bx)$ from the discrepancies between the measured acoustic lengths across the object and across a reference medium (water), 
i.e.\ to find $\bar{n}_{\text{object}}(\bx)$ from $ \Delta \bar{L} = c_w \Delta \bar{T}$, where $\Delta \bar{T} = \bar{T}_{\text{object}} -\bar{T}_w $ and $\bar{T}_w$ are the times-of-flight through water with known sound speed $\bar{c}_w$. 
This can be posed as a nonlinear minimisation problem of the form
\begin{align} 
\label{eq:nonlinear_minimisation_for_n}
\bar{n}_* = \argmin_{\bar{n}}   \   E \big[ \bar{n}  \big]     
= \argmin_{\bar{n}}    \   \|  A \big[\bar{n} \big]  -   A \big[  \bar{n}_w  \big]    -   \Delta \bar{L}   \|_2^2,
\end{align}
where the `object' subscript has been dropped and from now on is implied. This nonlinear problem can be solved as a sequence of linearised problems. A linearisation of the forward operator $A$ about a fixed $\bar{n}_q$, the q-th update for the refractive index field, gives the approximations
\begin{align} 
\label{eq:linearised_forward_operator}
\begin{split}
A \big [ \bar{n} \big]  & \approx A \big [ \bar{n}_q  \big]  +  J \big[  \bar{n}_q] \big( \bar{n}- \bar{n}_q \big)  \\
A \big [ \bar{n}_w  \big]  & \approx A \big [ \bar{n}_q  \big]   +  J \big[  \bar{n}_q] \big( \bar{n}_w - \bar{n}_q \big)
\end{split}
\end{align}
Substituting Eq.\ \eqref{eq:linearised_forward_operator} into Eq.\ \eqref{eq:nonlinear_minimisation_for_n} gives the \textit{q}-th linearised minimisation
\begin{align} 
\label{eq:qth_linearised_minimisation}
\bar{n}_{q+1} = \argmin_{   \bar{n} }   \  \tilde{E}_q  \big[    \bar{n}    \big]  
= \argmin_{ \bar{n} }     \   \|  J \big[ \bar{n}_q \big]   \Delta \bar{n}-    \Delta \bar{L}  \|_2^2,
\end{align}
where  $\Delta \bar{n}  = \bar{n} - \bar{n}_w $ with $\bar{n}_w =1 $ for all grid points.
Each linearised subproblem Eq.\ \eqref{eq:qth_linearised_minimisation} is solved by setting $\nabla \tilde{E}_q  \big[   \bar{n}  \big]= 0$, where the operator $\nabla \left(\cdot \right)$ denotes the gradient. (By convention, the first iteration is $q=0$.) 
As these are least squares problems, they lead to normal equations, which here are solved using a steepest descent algorithm with a fixed step length $1$, and terminated after a fixed number of iterations. The sequence of linearised subproblems is recursively repeated until 
\begin{align} 
\label{eq:stopping_criterion}
1- \tilde{E}\big[ \bar{n}_{q+1} \big]/  \tilde{E}\big[  \bar{n}_q \big] < \varepsilon_n,
\end{align} 
where $\varepsilon_n$ is a user-defined stopping threshold. 

\begin{remark}
Early-stopping of the steepest descent iterations in the solution of the subproblems can act implicitly as a regulariser, but an explicit regularisation term was not included in the objective function Eq.\ \eqref{eq:nonlinear_minimisation_for_n} so that the reconstructed image depends only on the proposed ray-based reconstruction algorithm and not on the value of a regularisation parameter. See, for example, \cite{Li1,Ali} for the application of regularisation to the UST inverse problem.
\end{remark}

\section{Ray tracing}   
\label{sec:Ray_tracing}
Here, the method used for finding the rays $f_{ (n,e,p)} (\bx) = 0$ (see Definition \ref{def:ray}) is described.

\subsection{Ray equation} 
\label{subsec:Ray_equation}
A general form of the acoustic wave equation in free space is
\begin{align}     
\label{eq:wave_equation}
u_{tt}(\bx,t)  - c(\bx)^2 u_{\bx\bx}(\bx,t) = 0, \ \bx \in \mathbb{R}^3, \ t > 0, 
\end{align}
where $u$ is the acoustic pressure, and $t$ time.
When $c$ is constant, a harmonic wave solution of Eq.\ \eqref{eq:wave_equation} can be written as $u(\bx,t) = U(\bx) \exp(i(\boldsymbol{k} \cdot \bx - \omega t)), $ where $U$ is the amplitude, $\omega$ is the angular frequency, and $ \boldsymbol{k}$ is a wavevector describing the direction of propagation. $U$, $\omega$ and $k$ are specified by the initial conditions \cite{Runborg}.
For a spatially varying $c(\bx)$, this can be modified, under a high frequency approximation, into the form \cite{Lay,Yang}
\begin{align}     
\label{eq:u_ansatz}
u(\bx,t) \approx U(\bx) e^{i\omega(W(\bx)/ c_0 - t )},
\end{align}
where $\boldsymbol{k} \cdot \bx $ has been replaced by  $(\omega/ c_0 ) W(\bx)$, the accumulated phase, and $c_0$ is a reference sound speed. For the high frequency approximation to hold, the fractional change in sound speed gradient over the longest wavelength involved in the problem must be small compared to the sound speed \cite{Lay}.
Now, substituting Eq.\ \eqref{eq:u_ansatz} into Eq.\ \eqref{eq:wave_equation} gives the two equations
\begin{align}  
\label{eq:two_ray_equations}
\nabla^2 U - \frac{\omega^2}{c_0^2}U\nabla W \cdot \nabla W  = - \frac{\omega^2} {c^2(\bx)} U, \quad 
2 \nabla W \cdot \nabla U + U \nabla^2 W = 0.
\end{align}
For the left-hand-side equation, neglecting the first term because of the same high frequency approximation gives \cite{Lay,Yang}
\begin{align}
\label{eq:eikonal}
\nabla W \cdot \nabla W  = n^2(\bx).
\end{align}
This is known as the \textit{eikonal equation}.
The surfaces on which $W(\bx)$ is constant are surfaces of constant phase called \textit{wavefronts}, and the lines which are always normal to these wavefronts are \textit{rays}. 
The acoustic energy propagates along these rays in the direction defined by the unit vector $d\bx_{ray}/ds$, where \cite{Anderson2}
\begin{align}
\frac{d\bx_{ray}}{ds}= \frac{ \nabla  W(\bx)} {n(\bx)}.
\end{align}
As $d/ds(n \, d\bx_{ray}/ds) = d/ds(\nabla W) = 
\nabla ( dW/ds) = \nabla n$, this leads directly to another form of the eikonal equation which here is called the \textit{ray equation}:
\begin{align}
\label{eq:ray_equation}
\frac{d}{ds} \left( n \frac{d\bx_{ray}(s)}{ds} \right) = \nabla n. 
\end{align}
In addition, the right-hand-side equation in Eq.\ \eqref{eq:two_ray_equations} is called \textit{transport equation}, and can be used for computation of the amplitude of the field along the ray \cite{Rawlinson2}, although it is not used in this study.

\subsection{Fermat's principle}
\label{subsec:Fermats_principle}
The ray equation can also be found using Fermat's principle \cite{Pierce1981, Holm}, which states that the path between two points taken by a ray makes the acoustic length stationary under variations in a family of nearby paths. As above, let $s$ denote the distance along a ray, and $\bx(s) \in \mathbb{R}^3$ and $ (d\bx/ds)(s) \in S^2$, denote the ray's position vector and (unit) direction vector, respectively. (The subscript on $\bx_{ray}(s)$ has been dropped for this section for conciseness.)

\begin{definition} 
\label{def:acoustic_length}
The acoustic length of a path $\bx(s)$ taken by a ray in passing from point $p_1$ to $p_2$ is defined by
\begin{align}
L_{(p_1,p_2)} = \int_{p_1}^{p_2} n\big(\bx(s)\big) ds,
\end{align}
where $ds = |d\bx(s)|$ is an infinitesimal distance along the ray (see Eq.\ \eqref{eq:forward_operator}). 
\end{definition}

\begin{definition} 
\label{def:family_of_paths}
A family of $C^2$ continuous paths $\bx(s, \epsilon) \in \mathbb{R}^3 $,  which smoothly depends on $\epsilon$, gives a set of smooth transformations of the ray path $\bx(s)$ in an interval including $\epsilon=0$, and satisfies \cite{Holm}
\begin{align} 
\label{eq:end_matching_conditions}
\bx(s_{p_1}, \epsilon) = \bx(s_{p_1}) , \  \bx(s_{p_2}, \epsilon) = \bx(s_{p_2}).
\end{align}
An infinitesimal variation of the path $ \bx(s)$ is defined in the form \cite{Holm}
\begin{align}
\delta \bx(s):= \frac{d}{d \epsilon} \Big|_{\epsilon = 0} \bx( s , \epsilon ),
\end{align}
where, from Eq.\ \eqref{eq:end_matching_conditions}, $\delta \bx(s_{p_1}) = \delta \bx(s_{p_2}) =0$.
\end{definition}

\begin{definition}  
\label{def:stationary_acoustic_length}
Using Definitions \ref{def:acoustic_length} and \ref{def:family_of_paths}, together with the fact that that $d \bx / ds $ is a unit vector, the stationary acoustic length under variation in a family of nearby paths satisfies \cite{Holm}
\begin{align} 
\label{eq:stationary_acoustic_length}
\delta L_{(p_1,p_2)} =  \delta   \bigg[ \int_{s_{p_1}}^{s_{p_2}}  n\Big(\bx(s)\Big)  \left( \frac{d \bx}{ds} \cdot  \frac{d \bx}{ds}  \right)^{1/2}  ds  \bigg] =0.
\end{align}
\end{definition}

\begin{lemma}
Using Definition \ref{def:acoustic_length} and the boundary conditions in Eq.\ \eqref{eq:end_matching_conditions}, the stationary of the optical length defined in Eq.\ \eqref{eq:stationary_acoustic_length} yields a path $\bx(s)$ that satisfies Eq.\ \eqref{eq:ray_equation}.
\begin{proof} 
Consider the integrand in Eq.\ \eqref{eq:stationary_acoustic_length} as a Lagrangian function in the form  $F(\bx, \bx', s ) = n  (\bx)  (  \bx'  \cdot  \bx' )^{1/2} $, where $ \bx' = d\bx/ ds $ has been used for brevity. Using Definition \ref{def:family_of_paths}, a perturbation of the first order, $\delta \bx$, is applied to the path $\bx(s)$ to give
\begin{align}
\begin{split}
\delta L_{(p_1,p_2)} & = \int_{s_{p_1}}^{s_{p_2}}  \Big[ F\big(\bx + \delta \bx  ,   \bx' +   \delta \bx' \big) - F \big(     \bx,  \bx' \big)  \Big] ds\\
&=  \int_{s_{p_1}}^{s_{p_2}}    \Big[ \delta \bx  \cdot  \frac{\partial F }{\partial \bx } + \delta \bx' \cdot  \frac{\partial F }{\partial \bx'}   \Big] ds.
\end{split}
\end{align}
Now, applying an integration by parts to the second term in the integrand, together with the boundary conditions in Definition \ref{def:family_of_paths}, gives  
\begin{align}
\delta L_{(p_1,p_2)} = \int_{s_{p_1}}^{s_{p_2}}   \delta \bx  \cdot  \Big(  \frac{\partial F }{\partial \bx } - \frac{d}{ds} \frac{\partial F}{\partial \bx'}  \Big)  ds.
\end{align}
The proof is straightforwardly completed by substituting $F$ into the above equation and setting the first order change $\delta L_{(p_1,p_2)} $ to zero for all perturbations $ \epsilon $.
\end{proof}
\end{lemma}

\subsection{Numerical ray tracing}  
\label{subsec:Numerical_ray_tracing}
In this section, two numerical ray-tracing algorithms are derived. 

\subsubsection{Dual-update algorithm}
Consider a ray starting from an emission point $\bx = \bx_{\bar{e}}$. A Taylor series expansion of the ray's position vector $\bx(s)$ can be written
\begin{align} 
\label{eq:TaylorSeries_x}
\bx(s+\Delta s) = \bx(s)+ \frac{d\bx}{ds} \bigg|_{s} \Delta s + \frac{1}{2} \frac{d^2\bx}{ds^2}\bigg|_{s} \Delta s^2 + O(\Delta s^3),
\end{align}
where $d\bx/ds|_s$ and $d^2\bx/ds^2|_s$ denote the first-order and second-order derivatives of the position vector at point $s$, respectively.
The second derivative can be found by expanding Eq.\ \eqref{eq:ray_equation} to get,
\begin{align}
\label{eq:ray_equation_expanded}
\frac{d^2 \bx}{ds^2}\bigg|_s = \frac{1}{n} \bigg[ \nabla n - \left( \nabla n \cdot \frac{d\bx}{ds}\bigg|_s  \right) \frac{d\bx}{ds}\bigg|_s \Bigg],
\end{align}
which is orthogonal to $d\bx/ds|_s$. For the first step, the first-order derivative, the ray tangent, is given by the user but for subsequent steps it needs to be found. A Taylor series expansion for $d\bx/ds$ can be written as 
\begin{align}
\label{eq:TaylorSeries_d}
\frac{d\bx}{ds}\bigg|_{s+\Delta s} = \frac{d\bx}{ds}\bigg|_s + \frac{d^2\bx}{ds^2}\bigg|_s \Delta s + O(\Delta s^2).
\end{align}
The two Taylor series, Eqs.\ Eq.\ \eqref{eq:TaylorSeries_x} and Eq.\ \eqref{eq:TaylorSeries_d} can be combined with
Eq.\ \eqref{eq:ray_equation_expanded} into a dual-update iterative scheme to calculate the next step along the ray $\bx(s+\Delta s)$ from the current point $\bx(s)$. This is written below as Algorithm \ref{alg:ray_tracing_dual_update}. The normalisation of the update vector in line 6 is included because it makes numerical integration along the ray easier if the ray is defined at evenly spaced points. The normalisation of the ray tangent vector, $\boldsymbol{d}$, in line 9 is there to ensure that it, like $d\bx/ds$ that it is an approximation to, is a unit vector.
In numerical tests in which $n$ and $\nabla n$ are known exactly, the error in the path of the ray as calculated by Algorithm \ref{alg:ray_tracing_dual_update} converges at a rate of $O(\Delta s^2)$, see Sec.\ \ref{subsubsec:Acoustic-length_convergence}. In many practical scenarios, such as in the inverse problem considered in this paper, the error in the ray path will, however, be dominated by error in the approximations
of $n$ and $\nabla n$ (Sec.\ \ref{subsec:Grid-to-ray_interpolation}). Furthermore, it is often the acoustic length - the integral of $n$ along the ray - that is the salient quantity, not the ray path \textit{per se}, and that will include errors due to the numerical approximation of the path integral as well. It can therefore be beneficial to trade a reduction in the formal rate of convergence of an algorithm for an increase in its computational efficiency, particularly when the former has little detrimental effect on the accuracy in the overall problem and the latter can significantly reduce the computational burden. With this in mind, a second ray-tracing algorithm was derived.
\begin{algorithm}
    \caption{Ray tracing: dual-update approach}
    \label{alg:ray_tracing_dual_update}
    \begin{algorithmic}[1]
        \State \textbf{input:} $\bx_{\bar{e}} \in \mathbb{R}^3$, $\bd_{\bar{e}} \in S^2$
            \Comment{Input initial ray position and tangent vectors}
        \State \textbf{initialise:} $\bx (0) = \bx_{\bar{e}}$, $\boldsymbol{d}(0) = \bd_{\bar{e}}$     \Comment{Set initial ray position and and tangent vectors}
        \While { $\bx(s) \ \text{is inside} \ \Omega$ } 
            \Comment{Run the loop while the ray remains in the domain}
            \State $\boldsymbol{h} = (\nabla n - \left( \nabla n \cdot \boldsymbol{d}  \right) \boldsymbol{d}) / n$ 
                \Comment{Compute the second derivative from Eq.\ \eqref{eq:ray_equation_expanded}}
            \State $\boldsymbol{g} = \boldsymbol{d} + \boldsymbol{h} \Delta s/2$
                \Comment{Calculate the ray position update direction}
            \State $\boldsymbol{g} \leftarrow \boldsymbol{g}/ \lvert \boldsymbol{g} \rvert$        
                \Comment{Make the update direction a unit vector}
            \State $\bx \leftarrow \bx + \boldsymbol{g} \Delta s$ 
                \Comment{Update the ray position vector}
            \State $\bd \leftarrow \bd + \boldsymbol{h}  \Delta s$ 
                \Comment{Update the ray tangent vector using Eq.\ \eqref{eq:TaylorSeries_d}}
            \State $\bd \leftarrow \bd / \lvert \bd \rvert$ 
                \Comment{Make the tangent vector a unit vector}
        \EndWhile
    \end{algorithmic}
\end{algorithm}

\subsubsection{Mixed-step algorithm}
The ray tracing algorithm derived in this section and used in the examples below, sometimes known as the mixed-step algorithm, \cite{Johnson, Denis} is computationally more efficient than Algorithm 1, although converges more slowly with $\Delta s$. Starting from the Taylor series expansion for the ray position vector, Eq.\ \eqref{eq:TaylorSeries_x}, and replacing the first derivative with the central difference approximation
$d\bx/ds|_s = (\bx(s+\Delta s) - \bx(s - \Delta s) ) / (2 \Delta s)$, 
gives a result which can be rearranged into the form
\begin{align} 
\label{eq:update_step_A}
\bx(s+\Delta s) = \bx(s) + \bd|_s \Delta s + \frac{d^2\bx}{ds^2}\bigg|_{s} \Delta s^2,
\end{align}
where $\bd|_s \coloneqq ( \bx(s) - \bx(s-\Delta s) ) / \Delta s$ is a backward difference. Eq.\ \eqref{eq:update_step_A} can be rewritten as
\begin{align} 
\label{eq:update_step_B}
\bx(s+\Delta s) = \bx(s) + \bd|_{s+\Delta s} \Delta s,
\end{align}
where the ray-direction update is given by 
\begin{align} 
\label{eq:d_update}
\bd|_{s+\Delta s} = \bd|_s + \frac{d^2\bx}{ds^2}\bigg|_{s} \Delta s,
\end{align}
(see the Taylor series, Eq.\ \eqref{eq:TaylorSeries_d}). Note that the ray-direction update $\bd|_{s+\Delta s}$ must be normalised to ensure that Fermat's principle holds (see Eq.\ \eqref{eq:ray_equation_expanded}) and it gives steps along the ray of equal length $\Delta s$. Using Eq.\ \eqref{eq:update_step_B} to calculate the next position along the ray requires $\bd|_{s+\Delta s}$ to be calculated from Eq.\ \eqref{eq:d_update}, which raises the question of how to compute $\d^2\bx/ds^2$. This 
must be done using the ray equation, Eq.\ \eqref{eq:ray_equation_expanded},
as it is via this equation that the refractive index affects the ray path, but $\d\bx/ds |_s$ in Eq.\ \eqref{eq:ray_equation_expanded} is not known. In deriving Eq.\ \eqref{eq:update_step_A} from the Taylor series, $\d\bx/ds|_s$ was approximated using a central difference. However, here the backward difference $\bd|_s$ is preferred for reasons of computational efficiency (and for Fermat's principle to hold it must be normalised, as mentioned above). The use of a backward difference raises one issue, which is that it is necessary to know $\bx(s-\Delta s)$ at each step, and for the first step this is not known. This can be overcome by defining $\bd|_0$ as the user-defined initial ray tangent, and using Eq.\ \eqref{eq:TaylorSeries_x} and Eq.\ \eqref{eq:ray_equation_expanded} to compute $\bx(\Delta s)$ given $\bx(0)$. The resulting algorithm is shown in Algorithm \ref{alg:ray_tracing_mixed_step}. This algorithm can be coded more efficiently than Algorithm \ref{alg:ray_tracing_dual_update} as it contains fewer steps and fewer stored variables; the trade-off is a reduction in the convergence rate due to the first-order backward difference approximation used in the ray equation. However, as mentioned above, this is rarely a problem in practice as typically greater errors arise from the approximation of the refractive index gradient and the numerical path integration (described in Secs.\ \ref{subsec:Grid-to-ray_interpolation} and \ref{subsubsec:Jacobian_matrix} below).

\begin{algorithm}
    \caption{Ray tracing: mixed-step approach}
    \label{alg:ray_tracing_mixed_step}
    \begin{algorithmic}[1]
        \State \textbf{input:} $\bx_{\bar{e}} \in \mathbb{R}^3$, $\bd_{\bar{e}} \in S^2$
            \Comment{Input initial ray position and tangent vector}
        \State \textbf{initialise:} $\bx = \bx_{\bar{e}}$, $\boldsymbol{d} = \bd_{\bar{e}}$     
            \Comment{Set initial ray position and tangent vector}
        \State $\boldsymbol{h} = (\nabla n - \left( \nabla n \cdot \boldsymbol{d}  \right) \boldsymbol{d}) / n$ 
            \Comment{Compute the initial second derivative using Eq.\ \eqref{eq:ray_equation_expanded}}            
        \State $\bd \leftarrow \bd + \boldsymbol{h}
        \Delta s /2$ 
            \Comment{Update the ray direction based on Eq.\ \eqref{eq:TaylorSeries_x}}
        \State $\bd  \leftarrow \boldsymbol{d}/ \lvert \boldsymbol{d} \rvert$        
            \Comment{Make the update direction a unit vector}
        \State $\bx \leftarrow \bx + \bd \Delta s$ 
            \Comment{Update ray position vector}        
        \While { $\bx(s) \ \text{is inside} \ \Omega$ } 
            \Comment{Run the loop while the ray remains in the domain}
            \State $\boldsymbol{h} = (\nabla n - \left( \nabla n \cdot \boldsymbol{d}  \right) \boldsymbol{d}) / n$ 
                \Comment{Compute the second derivative using Eq.\ \eqref{eq:ray_equation_expanded}}
            \State $\bd \leftarrow \bd + \boldsymbol{h}
        \Delta s$
                \Comment{Update the ray direction using Eq.\ \eqref{eq:d_update}}  
            \State $\bd  \leftarrow \boldsymbol{d}/ \lvert \boldsymbol{d} \rvert$       
                \Comment{Make the update direction a unit vector}   
            \State $\bx \leftarrow \bx + \bd \Delta s$ 
                \Comment{Update the ray position vector}
        \EndWhile
    \end{algorithmic}
\end{algorithm}

\subsection{Grid-to-ray interpolation}  
\label{subsec:Grid-to-ray_interpolation}
As mentioned in Sec.\ \ref{subsec:Discrete_forward_operator}, the refractive index is defined on a grid of points. In contrast, the points along the ray computed using the algorithms above can lie anywhere in $\Omega$ and are not restricted to grid points. An interpolation map from the grid to the rays is therefore necessary for approximating $n$ and $\nabla n$ in the ray tracing algorithms, as well as for forming the Jacobian matrix $J\big[ \bar{n}_q  \big]$ (cf.\ Definition \ref{def:Jacobian}) used for solving the inverse problem in Eq.\ \eqref{eq:qth_linearised_minimisation}.

\subsubsection{Numerical approximation of refractive index.}
The refractive index field was discretised on a mesh of points $x_i$ with $i \in \{ 1,...,N_n \}$. For simplicity, a rectilinear grid was used with grid points indexed with the multi-index $i = \big( i^1 , i^2 , i^3  \big) \in \left\{ 1, ... , N_n^1 \right\}  \times \left\{ 1, ... , N_n^2 \right\}  \times  \left\{ 1, ... , N_n^3 \right\} $ with $N_n = \prod_{j=1}^3 N_n^j $ and an equal grid spacing $\Delta x$ along all Cartesian coordinates $j$. Also, $x_{i^j}$ is used to indicate the position of grid point $i$
along Cartesian coordinate $j$. (Recall that $x^j$ denotes the position in continuous Euclidean space along the Cartesian coordinate $j$.)

Let $\{ \phi_i( \bx),i=1,...,N_n\}$ denote a set of basis functions for which basis function $\phi_i( \bx)$ is related to grid point $i$. An arbitrary continuous scalar field $z(\bx)$ (this could be the refractive index or a component of its gradient) can be approximated as a linear combination of these basis functions
\begin{align}  
\label{eq:linear_combination}
z(\bx) \approx \hat{z}( \bx  ) = \sum_{i=1}^{N_n} \bar{z}_i \phi_i( \bx  ).
\end{align}
Here, trilinear basis functions defined on a regular grid were used, which take the form \cite{Denis}
\begin{align} 
\label{eq:trilinear_basis_functions}
\phi_i(\bx) = \prod_{j=1}^3 \left(1- |u_i^j(\bx)| \right),
\end{align}
where
\begin{align}  
\label{eq:trilinear_basis_functions2}
\begin{split}
u_i^j  (\bx) = 
\begin{cases} 
\left(x^j - x_{i^j} \right) / \Delta x,  & x_{(i-1)^j}  <  x^j <  x_{(i+1)^j}  \\
0,  &\text{otherwise}.
\end{cases}
\end{split}
\end{align}  
Here, $x_{(i\pm 1)^j} $ denotes the two grid points adjacent to the grid point $x_i$ along the Cartesian coordinate $j$. The basis function $\phi_i(\bx)$ has a pyramidal shape with the vertex on point $i$, and it vanishes on the neighboring grid points. 
Because this basis function is a polynomial of first order along each Cartesian coordinate $j$, the interpolated function $\hat{z}$ is continuous but its first derivative is not continuous \cite{Denis}. Therefore, for an approximation of the directional gradients of the refractive index, we follow \cite{Anderson2,Anderson3}, where the discretised directional gradients are first calculated from the values of field at the grid nodes using finite differences, and are then approximated for off-grid points using the same interpolation as for the field itself. (Using this approach, the need for different interpolation operators for the refractive index and each of the three gradient components is avoided.) It has been shown that this approach gives an approximation that is sufficiently accurate for weakly heterogeneous media \cite{Anderson2,Anderson3,Denis}, which is the case here. We will show in Section \ref{sec:Numerical_results} that this approach provides a good trade-off between accuracy and computational cost.

\subsubsection{Interpolation operator.}
This section describes the map used for interpolating from the discretised field on the grid, $\bar{z}$, to an arbitrary point along the ray using Eq.\ \eqref{eq:linear_combination}.

\begin{definition}
The positions of the first and last vertices of each cubic voxel $v$ along each Cartesian coordinate $j$ are given by $x_{i'^j}^v$, where $i' \in \left\{  0,1  \right\}$ denotes the first and last vertices. Any arbitrary point $\bx = \{x^j,j=1,2,3\}$ contained in $v$ satisfies $ x_{0^j}^v  \leqslant x^j < x_{1^j}^v$
for all Cartesian coordinates $j$.
The values of the discretised field $\bar{z}$ associated with the vertices of voxel $v$, containing $\bx$, are notated $\bar{z}^v$ and defined using the map
\begin{align}  
\label{eq:field_to_voxel_map}
\begin{split}
J^v &: \mathbb{R}^{N_n} \rightarrow \mathbb{R}^{8}\\
\bar{z}^v \big[ \bx  \big]  &= J^v \big[ \bx \big] \bar{z}.
\end{split}
\end{align}
\end{definition}

Correspondingly, the elements of $\bar{z}^v$ are written as
\begin{align}
\bar{z}^v = \left[\bar{z}_{(0,0,0)}^v,\bar{z}_{(0,0,1)}^v,\bar{z}_{(0,1,0)}^v,\bar{z}_{(0,1,1)}^v,\bar{z}_{(1,0,0)}^v,\bar{z}_{(1,0,1)}^v,\bar{z}_{(1,1,0)}^v,\bar{z}_{(1,1,1)}^v\right]^T.
\end{align}
We will also use
\begin{align} 
\label{eq:big_Q}
Q^v  \big[ \bx\big]= \left[1, \ l^{(v,1)}, \ l^{(v,2)}, \ l^{(v,3)}, \ l^{(v,1)}l^{(v,2)},
 l^{(v,2)}l^{(v,3)}, \ l^{(v,1)}l^{(v,3)}, \ l^{(v,1)} l^{(v,2)} l^{(v,3)}\right],
\end{align}
where the dependence of $l^{(v,j)}$ on $\bx$ is neglected for brevity, but is given by
\begin{align}   
\label{eq:big_Q2}
l^{(v,j)}(\bx) = \frac{ \bx - x_{0^j}^v  }{  x_{1^j}^v - x_{0^j}^v }= \frac{ \bx - x_{0^j}^v  }{  \Delta x }.
\end{align}

\begin{definition}
The trajectory of a ray-linking an emission point $\bar{e}$ to a reception point $\bar{r}$ is defined by the points $s_m \ m \in \left\{ 0,...,M_{(\bar{e},\bar{r})} \right\} $, where the initial point $s_0$ matches the emission point $\bar{e}$, the final point $s_{M_{(\bar{e},\bar{r})}}$ matches the reception point $\bar{r}$, and the number of the sampling points along the ray is $M_{(\bar{e},\bar{r})}+1$. 
For ray tracing algorithms that take equal sized steps along the ray, as is the case for Algorithms \ref{alg:ray_tracing_dual_update} and \ref{alg:ray_tracing_mixed_step}, the points $s_m$ must satisfy
\begin{align}  
\label{eq:ray_points_sm}
\begin{split}
s_m = 
\begin{cases}
m \Delta s,                                & m\in\left\{0,..., M_{(\bar{e},\bar{r})} -1 \right\}\\
\left( m-1 \right)  \Delta s + \Delta s',   & m= M_{(\bar{e},\bar{r})}.
\end{cases}
\end{split}
\end{align}
Also, the second line in Eq.\ \eqref{eq:ray_points_sm} is used in order to indicate that the last point of the ray must be matched to the reception point $\bar{r}$, and thus $\Delta s' = s_{M_{(\bar{e},\bar{r})}} - s_{M_{(\bar{e},\bar{r})-1}}$ with $ \Delta s'  \leqslant  \Delta s$. This is achieved using ray-linking, described in Sec. \ref{sec:Two-point_ray_tracing_ray_linking}.
\end{definition}

\begin{lemma}
Using the trilinear basis function, Eq.\ \eqref{eq:trilinear_basis_functions}, an interpolation map (operator) from a discretised field $\bar{z}$ to sampling points $s_m$ on a ray is defined as
\begin{align}   
\label{eq:field_to_ray_interpolation}
\begin{split}
& J_{(s_m;\bar{e},\bar{r})}^{int}:  \mathbb{R}^{N_n} \rightarrow \mathbb{R}^{M_{(\bar{e},\bar{r})}+1}\\
& \hat{z}\big[\bx_{(s_m;\bar{e},\bar{r})}\big] =  
J_{(s_m;\bar{e},\bar{r})}^{int} \bar{z}
= Q^v\big[ \bx_{(s_m;\bar{e},\bar{r})}\big] C J^v \big[ \bx_{(s_m;\bar{e},\bar{r})} \big] \bar{z},
\end{split}
\end{align}  
where, the matrix $C \in \mathbb{R}^{8 \times 8}$ is in the form
\begin{align}
\begin{split}
C= 
\begin{bmatrix}
 1 & 0 & 0 & 0 & 0 & 0 & 0 & 0 \\
-1 & 0 & 0 & 0 & 1 & 0 & 0 & 0 \\
-1 & 0 & 1 & 0 & 0 & 0 & 0 & 0 \\
-1 & 1 & 0 & 0 & 0 & 0 & 0 & 0 \\ 
 1 & 0 &-1 & 0 & -1& 0 & 1 & 0 \\
 1 &-1 &-1 & 1 & 0 & 0 & 0 & 0 \\
 1 &-1 & 0 & 0 & -1& 1 & 0 & 0 \\
-1 & 1 & 1 & -1&  1& -1& -1& 1	
\end{bmatrix}.
\end{split}
\end{align}
\begin{proof}
The proof is obtained by forming $\hat{z}$ in Eq.\ \eqref{eq:linear_combination} using Eq.\ \eqref{eq:trilinear_basis_functions} and Eq.\ \eqref{eq:trilinear_basis_functions2}, and then re-arranging into the 
formula Eq.\ \eqref{eq:field_to_ray_interpolation} using Eq.\ \eqref{eq:big_Q} and Eq.\ \eqref{eq:big_Q2}. 
\end{proof}
\end{lemma}

\begin{remark}
The interpolation operator $J_{(\bar{e},\bar{r})}^{int}$ is used for two purposes: for approximating $n$ and $\nabla n$ along the rays in the ray tracing algorithm above, and in constructing the Jacobian matrix (Definition \ref{def:Jacobian}), described below.
\end{remark}

\subsubsection{Jacobian matrix} 
\label{subsubsec:Jacobian_matrix}
This section follows the definition for the Jacobian matrix $J$ in Definition \ref{def:Jacobian}, and gives further details about its action on a perturbation field $\delta \bar{n}$. It is worth mentioning that while $J$ is dependent on  $\bar{n}$, it acts on $\Delta \bar{n}$ in our iterative inversion algorithm (see Eq.\ \eqref{eq:qth_linearised_minimisation}).
From Eq.\ \eqref{eq:field_to_ray_interpolation}, $J_{(\bar{e},\bar{r})}^{int}$ maps $\delta \bar{n}$ on the grid points to $\delta \hat{n}$ on the sample points along one ray, the one between the emission point $\bar{e}$ and reception point $\bar{r}$. Given $\delta \hat{n}$, the perturbation in the acoustic length along that ray can be calculated using the map
\begin{align}  
\label{eq:perturbation_in_acoustic_length_map}
\begin{split}
&J_{(\bar{e},\bar{r})}^l: \mathbb{R}^{M_{(\bar{e},\bar{r})}+1} \rightarrow \mathbb{R}   \\
& \delta \bar{L}_{(\bar{e},\bar{r})}  =  J_{(\bar{e},\bar{r})}^l\delta \hat{n} \\
&=\frac{1}{2} \delta \hat{n}  \big[  \bx_{(  0;\bar{e},\bar{r})}  \big]  \Delta s +
\sum_{m=1}^{M_{(\bar{e},\bar{r})}-2}  \delta  \hat{n}  \big[  \bx_{( m \Delta s;  \bar{e},\bar{r} )}   \big]    \Delta s  \\
&  +   \frac{1}{2} \delta \hat{n} \big[ \bx_{\big(  ( M_{(\bar{e},\bar{r})} -1 )  \Delta s;\bar{e},\bar{r}  \big)} \big] \big( \Delta s + \Delta s' \big)   \\
& + \frac{1}{2} \delta \hat{n}  \big[  \bx_{\big(  ( M_{(\bar{e},\bar{r})} -1 )  \Delta s + \Delta s'  ; \bar{e},\bar{r} \big)}   \big]  \Delta s', 
\end{split}
\end{align}
which is a numerical path integration of the $\delta \hat{n}$ along the ray using the trapezoidal rule. Using $J_{(\bar{e},\bar{r})}$ to denote the row of the full Jacobian $J$ corresponding to an emission point $\bar{e}$ and a reception point $\bar{r}$, and considering Definition \ref{def:Jacobian}, Eq.\ \eqref{eq:field_to_ray_interpolation} and Eq.\ \eqref{eq:perturbation_in_acoustic_length_map}, the action of $J_{(\bar{e},\bar{r})}$ on a field $\delta \bar{n}$ can be written as
\begin{align}
\label{eq:Jacobian_discrete}
\begin{split}
&J_{(\bar{e},\bar{r})}:  \mathbb{R}^{N_n}\rightarrow \mathbb{R}\\
& \delta \bar{L}_{(\bar{e},\bar{r})} 
= J_{(\bar{e},\bar{r})}\delta \bar{n}
= J_{(\bar{e},\bar{r})}^l J_{(\bar{e},\bar{r})}^{int}  \delta \bar{n}.
\end{split}
\end{align}

\section{Ray-linking (Two-point ray tracing)}   
\label{sec:Two-point_ray_tracing_ray_linking}
The ray tracing algorithms derived in Section \ref{sec:Ray_tracing} solve a path for the rays $f_{(n;\bar{e},p)}=0$, where $\bar{e} \in \mathbb{S}_1$ is the emission point on the hemispherical surface $\mathbb{S}_1$, and $p \in \mathbb{S}$ is the interception point of the ray with the surface $\mathbb{S}$ after traveling through the medium. From Definitions \ref{def:discrete_forward_operator} and \ref{def:Jacobian}, the forward operator and Jacobian matrix are given by the path integral of the refractive index along a set of rays that link the emission points $ \bar{e} $ to the reception points $\bar{r}$. These rays are defined using $f_{(n;\bar{e},\bar{r})}=0$. Accordingly, for each pair of points $\bar{e}$ and $\bar{r}$, one seeks to find the ray's path $f_{(n;\bar{e},p)}=0$ such that $p$ matches $\bar{r}$. Applying this condition leads to a boundary value problem called \textit{ray-linking}. This boundary value problem can be solved as a sequence of initial value problems by adjusting the initial direction of the ray  until the ray intercepts $\mathbb{S}$ sufficiently close to the reception point $\bar{r}$ \cite{Sambridge,Anderson2,Anderson3,Rawlinson1, Rawlinson2}. This approach fits into a class of methods for solving boundary value problems known as \textit{shooting methods}.

An alternative class of methods for two-point ray tracing is those based on \textit{ray bending}. These kinds of approaches follow directly the Fermat's principle, and are based on fixing the two end points of the ray at the required  emission and reception points, and perturbing the ray trajectory until the acoustic length along the ray becomes stationary under these perturbations \cite{Um,Koketsu,Xu}. In this study, we use the former approach, ray-linking, and calculate the ray path between each pair of emitters and receivers by solving the inverse problem of finding the initial ray direction given the reception point through successive implementations of the ray-tracing algorithm. 

\subsection{Ray-linking inverse problem}
\begin{figure}[htb]
    \centering
	{\includegraphics[width=0.4\textwidth]{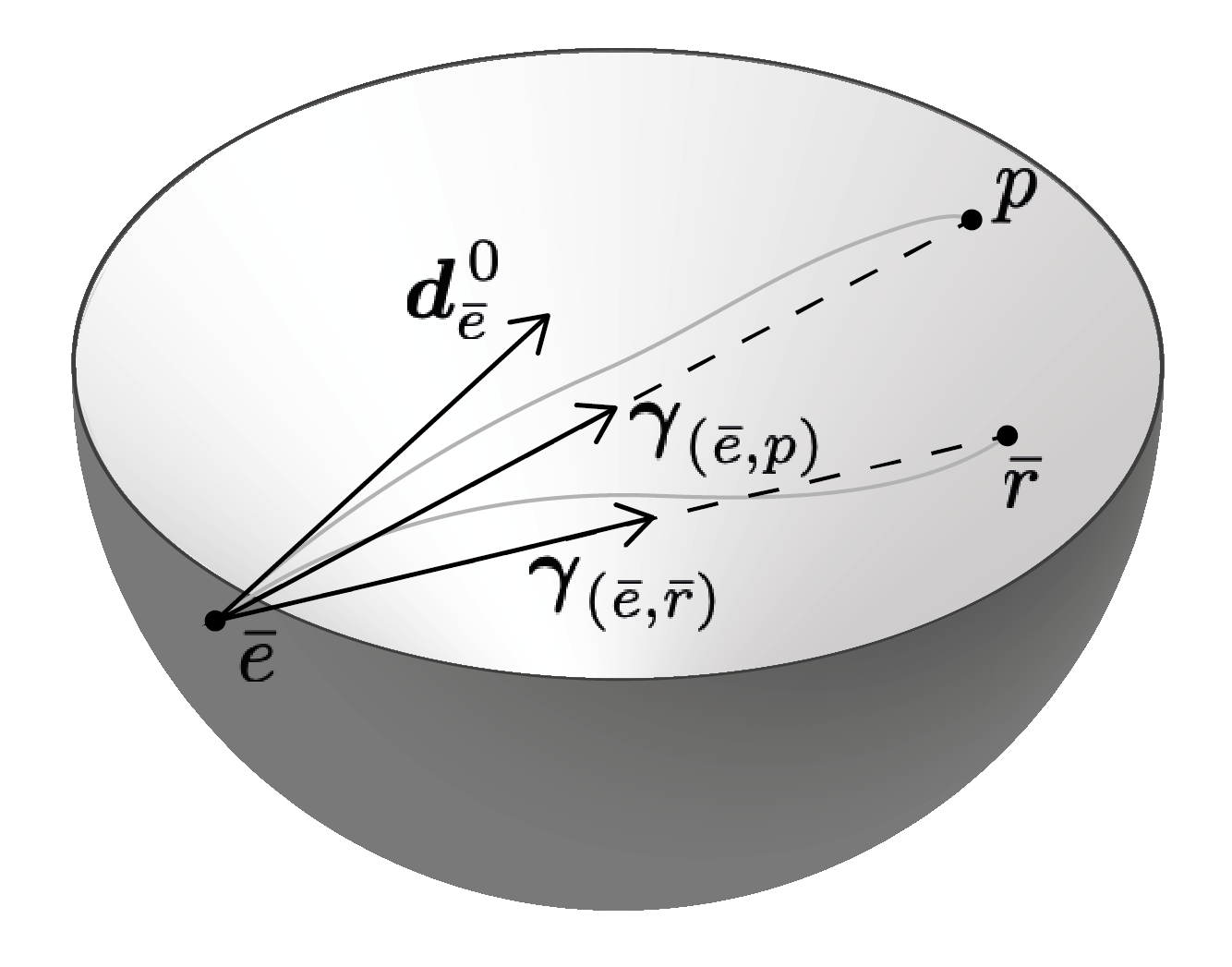}}
    \caption{The initial ray (grey line), with initial direction given by the unit vector $\bd_{\bar{e}}^0$, leaves the emission point $\bar{e} \in \mathbb{S}_1$ and arrives at the interception point $p \in \mathbb{S}$ (which is in a direction $\boldsymbol{\gamma}_{(\bar{e},p)}$ from $\bar{e}$). Following the ray-linking, the ray leaves $\bar{e}$ with a different initial direction and arrives at the desired reception point $\bar{r} \in \mathbb{S}_1$ (which is in a direction $\boldsymbol{\gamma}_{(\bar{e},\bar{r})}$ from $\bar{e}$). The ray-linking method works by iteratively adjusting the initial direction $\bd_{\bar{e}}$ until $\boldsymbol{\gamma}_{(\bar{e},p)} = \boldsymbol{\gamma}_{(\bar{e},\bar{r})}$.}
    \label{fig-raylinking}
\end{figure}

In the shooting method, the task of finding the ray that links the emission point $\bar{e}$ to the reception point $\bar{r}$ has been recast as the inverse problem of finding the initial ray-direction that results in the ray reaching the $\bar{r}$ (or close enough to it). This must be solved for all pairs of emission and reception points $(\bar{e}, \bar{r})$. Here, this is solved for each ray by iteratively adjusting the initial ray direction $\boldsymbol{d}_{\bar{e}}$ until the interception point $p$ matches the reception point $\bar{r}$. To facilitate this, we first parameterise the problem in a way that can be adapted to any convex detection surface.
Accordingly, the unknown initial direction of the ray $ \bd_{\bar{e}}=[\theta_{\bar{e}},\varphi_{\bar{e}}]^T \in \mathbb{R}^2 $ is written in terms of the azimuthal and polar angles . Then for each ray, the direction from the emission point $\bar{e}$ to the end point of the ray, the interception point $p$, is defined as a unit vector in spherical-coordinates centered on the emission point $\bar{e}$, i.e.,
\begin{align} 
\label{eq:gamma}
\begin{split}
 \mathbb{R}^2 &\rightarrow [0,2\pi)\times[0,\pi)\\
\boldsymbol{\gamma}_{(\bar{e},p)} \big[ \bd_{\bar{e}}  \big]   &  =  \boldsymbol{\gamma}_{(\bar{e},p)} \big[ \theta_{\bar{e}},     , \varphi_{\bar{e}}  \big]  
=\frac{\bx_p[\bd_{\bar{e}}] - \bx_{\bar{e}}}{\left|\bx_p[\bd_{\bar{e}}] - \bx_{\bar{e}}\right|},
\end{split}
\end{align}
where, from Algorithm \ref{alg:ray_tracing_mixed_step}, the dependence of $\bx_p$, and therefore $\boldsymbol{\gamma}_{(\bar{e},p)}$, on the unknown initial direction vector $\bd_{\bar{e}}$ is nonlinear. Here, $\theta_{\bar{e}} \in \big[ 0, 2 \pi \big)$ and $\varphi_{\bar{e}} \in  \big[ 0, \pi \big)$ are respectively the azimuthal and polar angles of the reception point $p$ with respect to the emission point $\bar{e}$. 
In the same way, the position of the reception point $\bar{r}$ is defined in terms of a unit direction vector $\boldsymbol{\gamma}_{(\bar{e},\bar{r})}$,
\begin{align}  
\boldsymbol{\gamma}_{(\bar{e},\bar{r})}=
\left( \theta_{(\bar{e},\bar{r})}, \varphi_{(\bar{e},\bar{r})} \right)^T
=\frac{\bx_{\bar{r}} - \bx_{\bar{e}}}{\left|\bx_{\bar{r}} - \bx_{\bar{e}}\right|}.
\end{align}
Using these definitions, the ray-linking problem for the pair of points $\bar{e}$ and $\bar{r}$ can be expressed either as a problem of solving for the root of the residual function $F : \mathbb{R}^2 \rightarrow  [-\pi,\pi)\times[-\pi,\pi) $:
\begin{align}  
\label{eq:ray_linking_root_finding}
F_{(\bar{e},\bar{r})} (\bd_{\bar{e}})  =  \boldsymbol{\gamma}_{(\bar{e},p)} \big[ \bd_{\bar{e}}  \big]  -   \boldsymbol{\gamma}_{(\bar{e},\bar{r})} = 0, 
\end{align}
or alternatively as a minimisation of the functional $\mathcal{E} : \mathbb{R}^2 \rightarrow \mathbb{R}$:
\begin{align} 
\label{eq:ray_linking_minimisation}
\argmin_{\bd_{\bar{e}}} \mathcal{E} (\bd_{\bar{e}}):= \frac{1}{2} \|     F_{(\bar{e},\bar{r})} (\bd_{\bar{e}}) \|_2^2.
\end{align}
The problem Eq.\ \eqref{eq:ray_linking_minimisation} is equivalent to solving for the root of the gradient of the residual function. (Note that the two elements of the residual function are wrapped to their corresponding intervals.)
A schematic for this ray-linking inverse problem is shown in Fig.~\ref{fig-raylinking}. Using the above parameterisation, this inverse problem can be thought of as an iterative adjustment of the initial direction $\bd_{\bar{e}}$ until the unit vector $\boldsymbol{\gamma}_{(\bar{e},p)}$ becomes aligned to the required unit vector $\boldsymbol{\gamma}_{(\bar{e},\bar{r})}$
(see Fig.~\ref{fig-raylinking}). In this way, the algorithms described below for solving the ray-linking inverse problem become independent from the geometry, and can be adapted to any arbitrary convex detection surface. 

\begin{remark}
A wide variety of standard solvers were applied to this problem but a significant portion of the rays always failed to link to the reception points. The reason for this is that the problem can become ill-posed in some cases. Adaptive-smoothing schemes are proposed to manage the ill-posedness, and are incorporated into three classical methods (Gauss-Newton, BFGS, and Broyden) in the three subsections below. The three methods are compared for accuracy as well as computational cost. 
\end{remark}

\subsection{Initial ray direction}
\label{subsec:Initial_ray_direction}
As described in Sec.\ \ref{subsec:Inverse_problem}, the refractive index map 
$\bar{n}_q$ is updated at every iteration $q$ of the UST reconstruction. The ray-linking problem must be solved for every pair of emitters and receivers $(\bar{e},\bar{r})$ for every iteration $q$, as the updated refractive index will change the ray trajectories. 
For each iteration $q>0$ the initial direction of the ray, $\bd_{(\bar{e},q)}^0$, is chosen to be the optimal ray from the previous iteration $\bd_{(\bar{e},q-1)}^*$. For the first iteration it is chosen to be the unit vector in the direction of $\bar{r}$, $\boldsymbol{\gamma}_{(\bar{e},\bar{r})}$:
\begin{align}   
\label{eq:initial_ray_direction}
\begin{split}
\bd_{(\bar{e},q)}^0 =
\begin{cases}
\bd_{(\bar{e},q-1)}^*, & q>0,\\
\boldsymbol{\gamma}_{(\bar{e},\bar{r})},  & q=0.
\end{cases}
\end{split}
\end{align}
For $q=0$, because $\bar{n}_0 = 1$, the Jacobian matrix $J \big[ \bar{n}_0 \big] $ in Eq.\ \eqref{eq:qth_linearised_minimisation} is also formed using 
$\boldsymbol{\gamma}_{(\bar{e},\bar{r})}$. As shown in the numerical results in Sec.\ \ref{sec:Numerical_results}, this choice provides very good initial guesses for the ray-linking problems. For the case of multiple linking paths, which will occur if the solution of the ray-linking problem is non-unique \cite{Anderson2}, this choice leads to the shortest path. 
An alternative way to choose the initial guesses is to shoot many initial rays from the emission point $\bar{e}$ then choose, for each reception point $\bar{r}$, the closest ray as the initial guess. However, our experience was that this approach worked less well than the proposed approach above, perhaps due to the existence of multiple linking paths.

\subsection{Ray-linking using functional minimisation: Damped Gauss-Newton}
\label{subsec:Gauss-Newton}
In the geophysics literature, a popular approach for ray-linking is solving the minimisation problem Eq.\ \eqref{eq:ray_linking_minimisation} based on an iterative linearisation of the functional $\mathcal{E}$ using finite differences, and solving a sequence of linearised minimization problems using Newton-type methods \cite{Julian,Sambridge,Rawlinson1,Rawlinson2}. Correspondingly, the \textit{k}-th linearisation of  $\boldsymbol{\gamma}_{(\bar{e},p)} $ around $\bd_{\bar{e}}^k $ gives
an approximation 
\begin{align} 
\label{eq:line_linearisation}
\boldsymbol{\gamma}_{(\bar{e},p)}   \big[ \bd_{\bar{e}} \big]   \approx  \boldsymbol{\gamma}_{(\bar{e},p)}  \big[ \bd_{\bar{e}}^k \big]   + \mathcal{J}  \big[ \bd_{\bar{e}}^k   \big]   \left( \bd_{\bar{e}} - \bd_{\bar{e}}^k \right).
\end{align}
Here, the superscript $k$ denotes the iteration number (the number of the linearised subproblem), and it is assumed that $ \bd_{\bar{e}}$ is sufficiently close to $ \bd_{\bar{e}}^k $. Also, $\mathcal{J} \big[ \bd_{\bar{e}}^k   \big] \in \mathbb{R}^{ 2 \times 2}$ denotes the partial derivative (Jacobian) matrix at $\bd_{\bar{e}}^k$, and is calculated using finite differences. ($\mathcal{J}$ should not be confused with its counterpart $J$ that stands for the Jacobian matrix (Definition \ref{def:Jacobian}) for the UST inverse problem .) Using Eq.\ \eqref{eq:line_linearisation}, the minimisation problem Eq.\ \eqref{eq:ray_linking_minimisation} can be modified into the form
\begin{align}  
\label{eq:direction_update_minimisation}
\boldsymbol{p}^k =  \argmin_{\boldsymbol{p}}    \   \tilde{\mathcal{E}}  \big[  \bd_{\bar{e}}^k  \big]  =   \argmin_{\boldsymbol{p}} \|  \mathcal{J}  \big[ \bd_{\bar{e}}^k   \big] \boldsymbol{p}   -  \Big ( \boldsymbol{\gamma}_{(\bar{e},\bar{r})} -  \boldsymbol{\gamma}_{(\bar{e},p)} \big[ \bd_{\bar{e}}^k \big]   \Big) \|_2^2,
\end{align} 
where $\tilde{\mathcal{E}}  \big[  \bd_{\bar{e}}^k  \big]$ is the \textit{k}-linearised functional, and $\boldsymbol{p}^k $ denotes the search direction. (The factor of a half has been dropped as it is not important.) The problem Eq.\ \eqref{eq:direction_update_minimisation} is solved for $\boldsymbol{p}^k$ by setting the gradient of the linearised functional to zero by forming the normal equations
\begin{align} 
\label{eq:normal_equations}
\nabla \tilde{\mathcal{E}}  \big[  \bd_{\bar{e}}^k  \big] = \mathcal{H}^k\boldsymbol{p}^k - \left( \mathcal{J}^k  \right)^T  \Big ( \boldsymbol{\gamma}_{(\bar{e},\bar{r})} -  \boldsymbol{\gamma}_{(\bar{e},p)} \big[ \bd_{\bar{e}}^k \big]   \Big)=0,
\end{align}
where $ {\mathcal{J}^k} := \mathcal{J} \big[ \bd_{\bar{e}}^k \big] $, and $ \mathcal{H}^k = \left( {\mathcal{J}^k} \right) ^T  \mathcal{J}^k $ is an approximation of the Hessian matrix in which the term of second order derivatives is neglected.

\subsubsection{Partial derivative (Jacobian) matrix}
The Jacobian matrix is in the form
\begin{align}  \label{jac}
\mathcal{J} \big[ \bd_{\bar{e}}^k \big]  =  \Big[  \frac{\partial \boldsymbol{\gamma}_{(\bar{e},p)}  }  {\partial  \theta_{\bar{e}} }   \big[ \theta_{\bar{e}}^k, \varphi_{\bar{e}}^k  \big],   \   \   \  \frac{\partial  \boldsymbol{\gamma}_{(\bar{e},p)}  }{ \partial \varphi_{\bar{e}} }  \big[ \theta_{\bar{e}}^k, \varphi_{\bar{e}}^k   \big]  \Big].
\end{align}
Here, the partial derivatives are calculated using finite differences
\begin{align}  
\label{eq:Jacobian_finite_difference}
\begin{split}
\frac{\partial \boldsymbol{\gamma}_{(\bar{e},p)} }{ \partial \theta_{\bar{e}} } \big[ \theta_{\bar{e}}^k, \varphi_{\bar{e}}^k  \big] &= \frac{\boldsymbol{\gamma}_{(\bar{e},p)} \big[ \theta_{\bar{e}}^k + \Delta \theta_{\bar{e}}^k , \varphi_{\bar{e}}^k  \big] - \boldsymbol{\gamma}_{(\bar{e},p)}  \big[ \theta_{\bar{e}}^k, \varphi_{\bar{e}}^k \big] }{ \Delta \theta_{\bar{e}}^k }\\
\frac{\partial \boldsymbol{\gamma}_{(\bar{e},p)} }{ \partial \varphi_{\bar{e}} } \big[ \theta_{\bar{e}}^k, \varphi_{\bar{e}}^k  \big] &= \frac{\boldsymbol{\gamma}_{(\bar{e},p)} \big[ \theta_{\bar{e}}^k  , \varphi_{\bar{e}}^k + \Delta \varphi_{\bar{e}}^k \big] - \boldsymbol{\gamma}_{(\bar{e},p)}  \big[ \theta_{\bar{e}}^k, \varphi_{\bar{e}}^k \big] }{ \Delta \varphi_{\bar{e}}^k },
\end{split}
\end{align}
where $\Delta \theta_{\bar{e}}^k$ and $\Delta \varphi_{\bar{e}}^k$ are perturbations enforced on the two elements of $\bd_{\bar{e}}^k$. How these perturbations are chosen is important. It has been shown that the inverse problem of ray-linking is subject to instability or divergence because of its highly nonlinear nature, especially for 3D cases \cite{Yang,Rawlinson2}. One important factor that affects the convergence is the accuracy of the Jacobian matrix \cite{Rawlinson2}, which is affected by the perturbations to the finite differences. A small perturbation gives a more accurate Jacobian matrix when $\boldsymbol{\gamma}_{(\bar{e},p)} $ is differentiable with respect to the initial direction $\bd_{\bar{e}}$, but the Jacobian matrix can become singular or close to singular when $\boldsymbol{\gamma}_{(\bar{e},p)} $ is nondifferentiable with respect to $\bd_{\bar{e}}$ or very sensitive to changes in $\bd_{\bar{e}}$. This may occur, for example, in the presence of singularities, such as when the trajectory of a ray is tangent to inter-medium boundaries, or when a ray is perpendicular to $\nabla n$, and thus a small change in $\bd_{\bar{e}}$ will lead to a large change in the position of the interception point of the ray with the detection surface $\mathbb{S}_1$. 
This issue, together with the low-dimensional nature of the ray-linking inverse problem, motivates using an adaptive approach for calculation of the Jacobian matrix. 
To facilitate this, each iteration of the ray-linking algorithm starts with a scaled perturbation in the form
\begin{align}    
\label{eq:initial_perturbations}
\begin{split}
\Delta \theta_{\bar{e}}^k = \tau \ \text{sign} \left( \theta_{\bar{e}}^k \right) \text{max} \left(  \text{abs} ( \theta_{\bar{e}}^k )    ,   \frac{1}{2}  \|\bd_{\bar{e}}^k \|_1 \right),\\
\Delta \varphi_{\bar{e}}^k = \tau \ \text{sign} \left( \varphi_{\bar{e}}^k \right) \text{max} \left(  \text{abs} ( \varphi_{\bar{e}}^k )    ,   \frac{1}{2}  \|\bd_{\bar{e}}^k \|_1 \right),
\end{split}
\end{align}
where $\tau$ is a small scalar. The perturbation is then increased recursively by controlling $\tau$ until the singular values of $\mathcal{H}^k$, $\Lambda^k \in \mathbb{R}^2$, satisfy the two conditions
\begin{align} 
\label{eq:singular_value_conditions}
\text{max}(\Lambda^k)  / \text{min}(\Lambda^k) < \vartheta 
\quad
\text{and}
\quad
\text{min}(\Lambda^k)  >   \text{min}( \mid \nabla \tilde{\mathcal{E}} \big[  \bd_{\bar{e}}^k  \big]  \mid , \varsigma ).     
\end{align}
Here, $\vartheta$ and $\varsigma$ are large and small scalar values, respectively. The former condition enforces a bound on the ill-conditioning of the Hessian matrix, and the latter prevents very large search directions (see Section 4.1 in \cite{Murray}). 

This scheme, here referred to as \textit{adaptive smoothing}, is outlined in Algorithm \ref{alg:adaptive_smoothing_Gauss_Newton}. The parameters $\eta$ and $N_h$ are user-adjusted parameters, but our numerical experience shows that the optimal values are independent of the particular object being imaged.

\begin{algorithm}
	\caption {\textit{Adaptive smoothing} calculation of well-conditioned Hessian and Jacobian matrices, $\mathcal{J}^k$ and $\mathcal{H}^k$, for the \textit{damped Gauss-Newton} approach to the ray-linking problem}
	\label{alg:adaptive_smoothing_Gauss_Newton}
	\begin{algorithmic}[1] \label{hes}
		\State \textbf{input:} $N_h$
		    \Comment{Maximum number of iterations}
		\State \textbf{input:}  $\vartheta$, $\varsigma$
		    \Comment{Scalars in the conditions in Eq.\ \eqref{eq:singular_value_conditions}}		    
		\State \textbf{input:}  $\eta$
		    \Comment{Integer factor $>1$ by which $\tau$ is recursively increased}
		\State \textbf{initialise:} $\tau$
		    \Comment{Controls the perturbation, Eq.\ \eqref{eq:initial_perturbations}}
		\State \textbf{initialise:} $k_h = 0$
		    \Comment{Set the counter}
		\State calculate $ \mathcal{J}^k $ and $ \mathcal{H}^k $ using Eq.\ \eqref{jac}, Eq.\ \eqref{eq:Jacobian_finite_difference}, Eq.\ \eqref{eq:initial_perturbations} and Alg. \ref{alg:ray_tracing_mixed_step}
		    \Comment{Form Jacobian and Hessian}
		\State calculate the singular values $\Lambda^k$ of $ \mathcal{H}^k $
		\While {$\Lambda^k$ does not satisfy Eq.\ \eqref{eq:singular_value_conditions} and $k_h < N_h$}
		    \Comment{Iterate until Hessian is well-conditioned}
            \State   $\tau \leftarrow \eta  \  \tau$ 
                \Comment{Increase the perturbation size}
		    \State   calculate $ \mathcal{J}^k $ and $ \mathcal{H}^k $ using Eq.\ \eqref{jac}, Eq.\ \eqref{eq:Jacobian_finite_difference}, Eq.\ \eqref{eq:initial_perturbations} and Alg. \ref{alg:ray_tracing_mixed_step}
			    \Comment{Form Jacobian and Hessian}
		    \State calculate the singular values $\Lambda^k$ of $\mathcal{H}^k $
            \State $ k_h   \leftarrow k_h +1 $
                \Comment{Increment the counter}        
		    \EndWhile
		\State \textbf{output:}  $ \mathcal{J}^k $ and $ \mathcal{H}^k $
		    \Comment{Return well-conditioned Jacobian and Hessian matrices}
	\end{algorithmic}
\end{algorithm}

\subsubsection{Backtracking line search}
Having defined the Jacobian matrix $\mathcal{J}^k$ and Hessian matrix $\mathcal{H}^k$, Eq.\ \eqref{eq:normal_equations} can be solved for the search direction $\boldsymbol{p}^k$. 
Once this is done, a search is conducted along $\boldsymbol{p}^k$ for a point that gives a sufficient descent step for $\mathcal{E} \big[ \bd_{\bar{e}}^k  \big]$:
\begin{align}  
\label{eq:d_line_search}
\bd_{\bar{e}}^{k+1} = \bd_{\bar{e}}^k + \alpha^k \boldsymbol{p}^k, 
\end{align} 
where $\alpha^k$ is the step length. Choosing $\alpha^k$ is the \textit{damping process} of the Gauss-Newton method, and is done here using a backtracking line search technique. The step length starts at $\alpha=1$ and is reduced recursively by a factor $\beta \in (0,1) $ until the following \textit{Armijo} condition is satisfied:
\begin{align}  
\label{eq:Armijo_condition}
\mathcal{E} \left(\bd_{\bar{e}}^k +\alpha^k   \boldsymbol{p}^k\right) \leqslant \mathcal{E} (\bd_{\bar{e}}^k) +  \mu   \alpha^k  \boldsymbol{p}^k   \nabla \mathcal{E}\left(\bd_{\bar{e}}^k \right) ,
\end{align}
where $\mu \in (0, 0.1 \big]$ is a user-adjusted parameter. In addition, a limit was set on the maximum number of recursions in the backtracking loop. (The full Wolfe conditions also require a recursive calculation of $\nabla \mathcal{E} \left(\bd_{\bar{e}}^k +\alpha^k   \boldsymbol{p}^k\right)$ for each iteration $k$, which is costly, so only the Armijo rule was used here.)

Algorithm \ref{alg:adaptive_smoothing_Gauss_Newton} can now be used to compute the initial ray direction to ensure ray-linking from the emission point $\bar{e}$ to the reception point $\bar{r}$. The whole ray-linking procedure using this functional minimisation approach, Eq.\ \eqref{eq:ray_linking_minimisation}, is outlined in Algorithm \ref{alg:ray_linking_Gauss-Newton}. 
In this algorithm, $\bd_{(\bar{e},q)}^0$ and $\bd_{(\bar{e},q)}^*$ denote the initial guess and an optimal solution respectively for the initial direction of the ray for linearisation $q$ of the UST inverse problem (Sec.\ \ref{subsec:Inverse_problem}). 
The choice of initial direction $\bd_{(\bar{e},q)}^0$ is described above in Sec.\ \ref{subsec:Initial_ray_direction}.

\begin{algorithm}
	\caption{Ray-linking from emission point $\bar{e}$ to reception point $\bar{r}$ using a damped Gauss-Newton method (for linearisation $q$ of the UST inverse problem)}
	\label{alg:ray_linking_Gauss-Newton}
	\begin{algorithmic}[1]
		\State  \textbf{input:} $\boldsymbol{\gamma}_{(\bar{e},\bar{r})}$
		    \Comment{Unit vector pointing to $\bar{r}$ from $\bar{e}$}
        \State  \textbf{input:} $N_{\text{link}}$
            \Comment{Maximum number of iterations}
        \State  \textbf{input:} $\varepsilon_{\text{link}}$
            \Comment{Minimum error threshold}
		\State  \textbf{initialise:} $\bd_{\bar{e}}^k = \bd_{(\bar{e},q)}^0$
		    \Comment{Set the initial ray direction}
		\State  \textbf{initialise:} $k=0$
		    \Comment{Set the counter}
		\State  calculate $\mathcal{E}(\bd_{\bar{e}}^k)$ using Eq.\ \eqref{eq:ray_linking_minimisation}
		    \Comment{Evaluate the error functional}
		\While {$  \mathcal{E} \left( \bd_{\bar{e}}^k   \right)  > \varepsilon_\text{link}$ \textbf{and} $k < N_\text{link} $}
		    \Comment{Iterate initial ray direction until $\boldsymbol{\gamma}_{(\bar{e},p)} = \boldsymbol{\gamma}_{(\bar{e},\bar{r})}$ }
            \State  calculate $ \mathcal{J}^k$ and $ \mathcal{H}^k$ using Algorithm \ref{alg:adaptive_smoothing_Gauss_Newton}
                \Comment{Form Jacobian and Hessian matrices}
	        \State  calculate $\boldsymbol{p}^k$ using Eq.\ \eqref{eq:normal_equations}
	            \Comment{Find descent direction}
            \State  choose $\alpha^k $ that satisfies Eq.\ \eqref{eq:Armijo_condition}
	            \Comment{Backtracking line search to find step size}
            \State  $\bd_{\bar{e}}^{k+1} \leftarrow \bd_{\bar{e}}^k  + \alpha^k \boldsymbol{p}^k $
                \Comment{Update initial ray direction}
            \State  calculate $\mathcal{E}(\bd_{\bar{e}}^{k+1})$ using Eq.\ \eqref{eq:ray_linking_minimisation}
		        \Comment{Evaluate the error functional}
            \State  $k \leftarrow k+1$ 
                \Comment{Increment the counter}
		\EndWhile
		\State \textbf{output:} $\bd_{(\bar{e},q)}^*$
		    \Comment{Return the optimal initial ray direction}
	\end{algorithmic}
\end{algorithm}

\subsection{Ray-linking using root-finding: Quasi-Newton with box constraints}
An analogous approach to solving Eq.\ \eqref{eq:ray_linking_minimisation} using the Gauss-Newton approach would be to solve the system of nonlinear equations Eq.\ \eqref{eq:ray_linking_root_finding} using the \textit{Newton-Raphson} method \cite{Galanti}. This approach gives a sequence of linear equations
\begin{align} 
\label{eq:normal_equations2}
\mathcal{J}^k\boldsymbol{p}^k  +F \big[ \bd_{\bar{e}}^k \big]=0,
\end{align}
where, compared to Eq.\ \eqref{eq:ray_linking_root_finding}, the subscripts $(\bar{e},\bar{r})$ for the function $F$ has been neglected for brevity.
The Newton-Raphson method, like the Gauss-Newton approach, requires frequent calculation of the Jacobian, which is computationally costly. Each iteration of a Newton-type method (or a trust region method \cite{Bellavia}), used for solving either Eq.\ \eqref{eq:ray_linking_root_finding} or Eq.\ \eqref{eq:ray_linking_minimisation}, requires at least two additional function evaluations per iteration to calculate the Jacobian. (Note that if a higher order finite difference scheme is used to increase the accuracy over formulae Eq.\ \eqref{eq:Jacobian_finite_difference}, the number of function evaluations will increase accordingly.) Fortunately, there is a class of approaches for solving nonlinear equations such as Eq.\ \eqref{eq:normal_equations2} - \textit{Quasi-Newton} methods - that have the great advantage of not requiring Jacobian calculations.
For nonlinear equations, Quasi-Newton methods suggest a replacement of the sequence of equations Eq.\ \eqref{eq:normal_equations2} by derivative-free linear equations of the form
\begin{align} 
\label{eq:normal_equations_Quasi-Newton}
\mathcal{B}^k\boldsymbol{p}^k  +F\big[ \bd_{\bar{e}}^k \big]=0, 
\end{align}
where $\mathcal{B}^k$ is a matrix that approximates $\mathcal{J}^k$ using the two last updates. There are different formulas for forming the approximate Jacobian $\mathcal{B}^k$. Here, projected variants of the BFGS formula and a Broyden-like scheme will be described, as they have been found to give good results.

\subsubsection{BFGS-like formula}
For updating the matrix $\mathcal{B}$, one way is to use a modified variant of the BFGS formula in the form
\begin{align}  
\label{eq:BFGS_update}
\mathcal{B}^{k+1} = \mathcal{B}^k + \tau^k \left( - \frac{\mathcal{B}^k s^k {s^k}^T   \mathcal{B}^k }{{s^k}^T \mathcal{B}^k s^k }+ \frac{y^k {y^k}^T}{{y^k}^T s^k} \right),
\end{align}
where $s^k = \bd_{\bar{e}}^{k+1} - \bd_{\bar{e}}^k$ and $y^k= F(\bd_{\bar{e}}^{k+1}) - F(\bd_{\bar{e}}^k)$. $\bd_{\bar{e}}^{k+1}$ is calculated using the update formula Eq.\ \eqref{eq:d_line_search} and then projected onto a set of box constraints, as described in Sec.\ \ref{subsubsec:box_constraints} below.
The two terms in the parentheses are two symmetric rank-one matrices, but their summation provides a rank-two update for $\mathcal{B}$. 
Setting the parameter $\tau^k \in \left\{0,1\right\}$ to $1$ for all $k$ gives the standard BFGS formula \cite{Gu,Li4}. $\mathcal{B}^k$, in Eq.\ \eqref{eq:normal_equations_Quasi-Newton}, needs to be invertible, so that we can solve for $\boldsymbol{p}^k$. For unconstrained optimisation, the Wolfe conditions may be used to ensure $\mathcal{B}^k$ is positive definite, and therefore invertible, but they are not applicable to the root-finding problem here \cite{Li4}. Modified BFGS approaches have been proposed that ensure positive definiteness \cite{Gu,Zhou} but these approaches require at least one more function evaluation per iteration than vanilla BFGS, and in any case are applicable only to symmetric nonlinear equations, so not relevant here. A BFGS trust-region method has also been proposed for nonlinear equations \cite{Yuan}, but it relies on the Jacobian matrix which is what we are trying to avoid.

Here the nonsingularity of the matrix $\mathcal{B}$ is ensured using adaptive smoothing, i.e.\ iterative adjustment of $\tau$ as described in Algorithm \ref{alg:adaptive_smoothing_BFGS} below.
For the  Broyden-like method below, Sec.\ \ref{subsubsec:Broyden-like}, the parameter $\tau$ has been included in previous theoretical work \cite{Powell,More,Li5}. For the BFGS-like scheme, to the best of our knowledge, this is not the case but in our experience the method works well numerically. (This may be because for the optimisation case, for which BFGS is more commonly used, the positive definiteness of the update of $\mathcal{B}$ can be ensured using the Wolfe conditions.) Indeed, it is shown in the results section below that this BFGS-like formula, when combined with a box constraint on the sequence of initial directions, is much more efficient for ray-linking than the Gauss-Newton method above, which relies on the costly calculation of the Jacobian matrix.

\begin{algorithm}
	\caption {\textit{Adaptive smoothing} calculation of the approximate Jacobian matrix, $\mathcal{B}^k$, for the \textit{BFGS-like} approach to the ray-linking problem}
	\label{alg:adaptive_smoothing_BFGS}
	\begin{algorithmic}[1]
		\State \textbf{input:} $\mathcal{B}^k$
		    \Comment{Latest Jacobian approximation}
		\State \textbf{input:} $N_h$
		    \Comment{Maximum number of iterations}
		\State \textbf{input:}  $\vartheta$, $\varsigma$
		    \Comment{Scalars in the conditions in Eq.\ \eqref{eq:singular_value_conditions}}		    
		\State \textbf{input:} $\eta \in (0,1)$
		    \Comment{Factor by which $\tau$ is recursively decreased}
		\State \textbf{initialise:} $\tau = 1$
		\Comment{Weights the update to $\mathcal{B}^k$, Eq.\ \eqref{eq:BFGS_update}}
		\State calculate the singular values $\Lambda^k$ of $ \mathcal{B}^k $		
		\While {Eq.\ \eqref{eq:singular_value_conditions} is not satisfied \textbf{and} $ k_h < N_h $}
		    \Comment{Iterate until Jacobian $\mathcal{B}$ is well-conditioned}
            \State   $\tau \leftarrow \eta  \  \tau$ 
                \Comment{Reduce the update weight}
		    \State   calculate $ \mathcal{B}^{k+1} $  using Eq.\ \eqref{eq:BFGS_update}
                \Comment{Update the Jacobian approximation}
		    \State   calculate the singular values $\Lambda^k$ of $\mathcal{B}^{k+1}$  
            \State    $ k_h   \leftarrow k_h +1 $
                \Comment{Increment the counter}
		\EndWhile
		\State   \textbf{output:}  $\mathcal{B}^{k+1}$
		    \Comment{Return the updated Jacobian approximation}
	\end{algorithmic}
\end{algorithm}

\subsubsection{Broyden-like formula}
\label{subsubsec:Broyden-like}
Another way to approximate the Jacobian matrix $\mathcal{B}^k$, which is popular for solving nonlinear systems, is to use a Broyden-like formula \cite{Powell,More,Li5,Li4}. Initially, $\mathcal{B}^0 = \mathcal{J}^0 $, where $\mathcal{J}^0$ is calculated using Eq.\ \eqref{eq:Jacobian_finite_difference}, with a large enough perturbation to ensure nonsingularity, then for $k\ge 1$ this approximation is updated using the formula 
\begin{align} 
\label{eq:Broyden_update}
\mathcal{B}^{k+1} = \mathcal{B}^k + \tau^k \frac{(y^k - \mathcal{B}^k s^k) s_k^T }{{s^k}^T s^k},
\end{align}
where $ | \tau^k-1 | \leqslant  \bar{\tau}$ with $\bar{\tau} \in (0,1)$ a fixed scalar \cite{Powell,Li5,Li4,More}. 
When $\tau^k=1$ for all $k$, \eqref{eq:Broyden_update} becomes \textit{Broyden's rank one} formula \cite{Broyden}, and in this case a nonsingular $\mathcal{B}^k$ does not necessarily ensure a nonsingular $\mathcal{B}^{k+1}$.
For our application, this can occur when a small change in the initial angle leads to a large change in the end point of the ray \cite{Yang,Rawlinson2}. 
To avoid this issue, \textit{adaptive smoothing} is used, i.e.\ the singular values $\Lambda$ of the matrix $\mathcal{B}$ are controlled using a scalar $\tau$ initialised as $1$, and gradually moved away up to a neighborhood of radius $\bar{\tau}$ until the conditions Eq.\ \eqref{eq:singular_value_conditions} are satisfied. (In the second condition the gradient is replaced by $\mathcal{E} \big[  \bd_{\bar{e}}^k  \big] $. See Section 4.1 in \cite{Murray}.) This is outlined in Algorithm \ref{alg:adaptive_smoothing_Broyden}. 
Note that adaptive smoothing via a recursive adjustment of $\tau$, as used in Algorithms \ref{alg:adaptive_smoothing_BFGS} and \ref{alg:adaptive_smoothing_Broyden}, is very cheap compared to the approach used in Algorithm \ref{alg:adaptive_smoothing_Gauss_Newton}, as it does not require any additional function evaluations.

\begin{algorithm}
	\caption {\textit{Adaptive smoothing} calculation of the approximate Jacobian matrix, $\mathcal{B}^k$, for the \textit{Broyden-like} approach to the ray-linking problem}
	\label{alg:adaptive_smoothing_Broyden}
	\begin{algorithmic}[1]
		\State \textbf{input:} $\mathcal{B}^k$
		    \Comment{Latest Jacobian approximation}
		\State \textbf{input:} $N_h$
		    \Comment{Maximum number of iterations}
		\State \textbf{input:}  $\vartheta$, $\varsigma$
		    \Comment{Scalars in the conditions in Eq.\ \eqref{eq:singular_value_conditions}}		    
		\State \textbf{input:} $\eta_c \in (0, \bar{\tau})$
		    \Comment{magnitude of increment to $\eta$}
		\State \textbf{input:} $\bar{\tau}$
            \Comment{Limits size of weight $\tau$}
		\State \textbf{initialise:} $\eta=0$
            \Comment{Sets initial size of weight update}
		\State \textbf{initialise:} $\eta_s=1$
            \Comment{Sign of weight update}
		\State \textbf{initialise:} $\tau = 1$
    		\Comment{Weights the update to $\mathcal{B}^k$, Eq.\ \eqref{eq:Broyden_update}}
		\State   calculate the singular values $\Lambda^k$ of $ \mathcal{B}^k $
        \While  {$\Lambda^k$ does not satisfy Eq.\ \eqref{eq:singular_value_conditions} 		\textbf{and} $|\tau-1| \leqslant \bar{\tau}$ \textbf{and} $k_h<N_h$}

            \Comment{Iterate until Jacobian $\mathcal{B}$ is well-conditioned}
            \If {$\eta_s>0$}
		        \State $\eta \leftarrow \eta + \eta_c$
		            \Comment{increase weight update magnitude}
		    \EndIf
		    \State $\eta_s \leftarrow - \eta_s $
		        \Comment{alternate weight update sign}
            \State   $\tau= 1 + \eta_s \eta  $
		        \Comment{calculate the weight}
		    \State   calculate $ \mathcal{B}^{k+1} $  using Eq.\ \eqref{eq:Broyden_update}
		        \Comment{Update the Jacobian approximation}
		    \State   calculate the singular values $\Lambda^k$ 
            \State    $ k_h   \leftarrow k_h +1 $
		        \Comment{Increment the counter}
		\EndWhile
		\State  \textbf{output:}  $\mathcal{B}^{k+1}$
		    \Comment{Return the updated Jacobian approximation}
	\end{algorithmic}
\end{algorithm}

\subsubsection{Box constraints}
\label{subsubsec:box_constraints}
The reason for choosing the Gauss-Newton method to solve the minimisation problem Eq.\ \eqref{eq:ray_linking_minimisation} in Section \ref{subsec:Gauss-Newton} is the fast convergence rate of Newton-type methods for unconstrained minimisation problems. 
Similarly, the convergence of quasi-Newton methods for solving Eq.\ \eqref{eq:ray_linking_minimisation} is well-established \cite{Li4}, but for nonlinear equations such as Eq.\ \eqref{eq:ray_linking_root_finding}, less progress has been made \cite{Li4,Zhou}, a significant challenge is the lack of line-search techniques for algorithms that avoid costly derivative calculations \cite{Li4}. A good review of derivative-free line search techniques is given in \cite{Li4} and the reader is also referred to \cite{Griewank,Li6,Cruz}. However, these algorithms are typically conservative, and the high number of function evaluations required per iteration risks losing the computational advantage of not calculating gradients. 
Here, to achieve higher efficiency, an undamped variant of the update formula Eq.\ \eqref{eq:d_line_search} was used, with $\alpha^k=1$ for all $k$. If the search direction $\boldsymbol{p}^k$ obtained from Eq.\ \eqref{eq:normal_equations_Quasi-Newton} is not a descent direction for $\mathcal{E}$ at $\bd_{\bar{e}}^k$, then the sequence ${\bd_{\bar{e}}^k,k=0,1,\ldots}$ is likely to diverge. If the update direction $\boldsymbol{p}^k$ \textit{is} a descent direction, then, with an undamped scheme, it is still possible that the update step is large enough for the sequence to diverge.
To ameliorate this, a box constraint
of the form $\Gamma = \left\{ \bd_{\bar{e}} \in \mathbb{R}^2 \  |  \ \boldsymbol{l} \leqslant  \bd_{\bar{e}} \leqslant \boldsymbol{u} \right\} $, was enforced, where $\boldsymbol{l} \in \mathbb{R}^2 $ and $\boldsymbol{u} \in \mathbb{R}^2 $ are specified lower and upper bounds, and the inequalities are component-wise. The bounds are chosen as angular intervals containing an initial guess. Since $\boldsymbol{p}^k$ does not guarantee that the update $\bd_{\bar{e}}^k + \boldsymbol{p}^k$ is a feasible point, $\bd_{\bar{e}}^k + \boldsymbol{p}^k$ is projected onto the set $\Gamma$. To do this, a vector $\psi$ is defined as
\begin{align}  
\label{eq:box_constraint1}
\begin{split}
\psi_i^k=
\begin{cases}
\zeta \left(\frac{\boldsymbol{l}_i- \bd_{(\bar{e},i)}^k }{\boldsymbol{p}_i^k}\right),   &\text{if} \  \bd_{(\bar{e},i)}^k+ \boldsymbol{p}_i^k < \boldsymbol{l}_i, \\
\zeta \left(\frac{\boldsymbol{u}_i- \bd_{(\bar{e},i)}^k }{\boldsymbol{p}_i^k}\right),   &\text{if}  \ \bd_{(\bar{e},i)}^k+ \boldsymbol{p}_i^k > \boldsymbol{u}_i, \\
1, & \text{otherwise,}
\end{cases}
\end{split}
\end{align}
where $\zeta \in (0,1)$ is a scalar, and $i \in \left\{ 1,2 \right\}$ denotes the components. The update direction is then modified using
\begin{align}
\label{eq:box_constraint2}
\bar{\boldsymbol{p}}_i^k= \text{sign}(\psi_i^k)\Big(\text{max}\big(\text{abs}(\psi_i^k), \kappa\big)\Big) \boldsymbol{p}_i^k,
\end{align}
where $\kappa$ is a very small scalar, used to prevent the direction becoming stuck at one or other of the limits.
Note that because this minimisation is low-dimensional (two dimensions), a violation of the box constraint for even one component means that $\boldsymbol{p}^k$ may not be a descent direction, or may be a descent direction towards a local minimum far from the initial guess. In these cases, the update is projected to the interior of the feasible set, rather than the active set, using the parameter $\zeta$, which is heuristically chosen to be 0.5.

\begin{algorithm}
	\caption{Ray-linking from emission point $\bar{e}$ to reception point $\bar{r}$ using a Quasi-Newton method (for linearisation $q$ of the UST inverse problem)}
	\label{alg:ray_linking_Quasi_Newton}
	\begin{algorithmic}[1]
		\State  \textbf{input:} $\boldsymbol{\gamma}_{(\bar{e},\bar{r})}$
		    \Comment{Unit vector pointing to $\bar{r}$ from $\bar{e}$}
	    \State  \textbf{input:} $\boldsymbol{l}$, $\boldsymbol{u}$
		    \Comment{Lower and upper bounds on the direction vector $\bd_{\bar{e}}^k$}		    
        \State  \textbf{input:} $N_{\text{link}}$
            \Comment{Maximum number of iterations}
        \State  \textbf{input:} $\varepsilon_{\text{link}}$
            \Comment{Minimum error threshold}
		\State  \textbf{initialise:} $\bd_{\bar{e}}^k = \bd_{(\bar{e},q)}^0$
		    \Comment{Set the initial ray direction}
		\State  \textbf{initialise:} $k=0$
		    \Comment{Set the counter}
		\State calculate $\mathcal{B}^0 = \mathcal{J}^0$ using Eq.\ \eqref{eq:Jacobian_finite_difference}
	        \Comment{Form the approximate Jacobian}
		\State  calculate $\mathcal{E}(\bd_{\bar{e}}^k)$ using Eq.\ \eqref{eq:ray_linking_minimisation}
		    \Comment{Evaluate the error functional}
		\While {$  \mathcal{E} \left( \bd_{\bar{e}}^k   \right)  > \varepsilon_\text{link}$ \textbf{and} $k < N_\text{link} $}
		    \Comment{Iterate initial ray direction until $\boldsymbol{\gamma}_{(\bar{e},p)} = \boldsymbol{\gamma}_{(\bar{e},\bar{r})}$ }
		 	\State calculate $\boldsymbol{p}^k$ using Eq.\ \eqref{eq:normal_equations_Quasi-Newton}
		 	    \Comment{Update to direction vector}
		    \State calculate $\bar{\boldsymbol{p}}^k$ using $\boldsymbol{l}$ and $\boldsymbol{u}$ using Eq.\ \eqref{eq:box_constraint1} and Eq.\ \eqref{eq:box_constraint2} 
		        \Comment{Box constraints}
            \State  update   $\bd_{\bar{e}}^{k+1} \leftarrow \bd_{\bar{e}}^k  +  \bar{\boldsymbol{p}}^k $
                \Comment{Apply direction vector update}
            \State calculate $s^k = \bd_{\bar{e}}^{k+1} -     \bd_{\bar{e}}^k$
                \Comment{Difference used in Jacobian-update formula}
            \State calculate $F(\bd_{\bar{e}}^{k+1})$ using Eq.\ \eqref{eq:ray_linking_root_finding}
                \Comment{Difference between $\boldsymbol{\gamma}_{(\bar{e},\bar{r})}$ and  $\boldsymbol{\gamma}_{(\bar{e},p)}$}
            \State calculate $y^k= F(\bd_{\bar{e}}^{k+1}) - F(\bd_{\bar{e}}^k)$
                \Comment{Difference used in Jacobian-update formula}
            \State  calculate $ \mathcal{B}^{k+1}$ using Alg.\ \ref{alg:adaptive_smoothing_BFGS} (BFGS) or Alg.\ \ref{alg:adaptive_smoothing_Broyden} (Broyden)
            
                \Comment{Update approximation Jacobian}         
            \State  calculate $\mathcal{E}(\bd_{\bar{e}}^{k+1})$ using Eq.\ \eqref{eq:ray_linking_minimisation}
		        \Comment{Evaluate the error functional}
            \State  $k \leftarrow k+1$ 
                \Comment{Increment the counter}
		\EndWhile
		\State \textbf{output:} $\bd_{(\bar{e},q)}^*$
		    \Comment{Return the optimal initial ray direction}
	\end{algorithmic}
\end{algorithm}

\section{Numerical results}  
\label{sec:Numerical_results}
Having laid out our approach to the UST inverse problem
of estimating the refractive index in Sec.\ \ref{subsec:Inverse_problem}, and, above, the details of the ray tracing and ray-linking algorithms that will be used to achieve this, this section will describe numerical experiments demonstrating the effectiveness of this approach to 3D UST.

First, in Sec.\ \ref{subsec:Numerical_validation_ray_tracing}, the accuracy of the ray tracing algorithms, Algorithms \ref{alg:ray_tracing_dual_update} and \ref{alg:ray_tracing_mixed_step}, will be demonstrated on both continuous (analytically-known gradient) and discretised refractive index maps based on the Maxwell fish-eye phantom.
In Sec.\ \ref{subsec:Numerical_validation_ray_tracing}
these algorithms are used in calculations of acoustic length (the integral of the refractive index along a ray), and it is shown that these calculations are dominated by the grid-to-ray interpolation error and therefore Algorithm 
\ref{alg:ray_tracing_mixed_step} is more efficient for solving the UST problem.

Section \ref{subsec:Simulation_UST_data} describes the more complex breast phantom that was used in the subsequent UST study, and how the measured time series were simulated. (The time-of-flight picking method used to calculate the first-arrival of the signals from the simulated time series is described in an Appendix.) Finally, in Sec.\ \ref{subsec:UST_image_reconstruction}, all the previous work is brought together to demonstrate how to solve the problem of estimating the refractive index (sound speed) in 3D ultrasound tomography of the breast. Our bent ray approach is compared to an approach using straight-rays, and the different ray-linking algorithms are compared with respect to accuracy and computational time. 

\subsection{Numerical validation of ray tracing algorithms}
\label{subsec:Numerical_validation_ray_tracing}

\subsubsection{Maxwell's fish-eye lens phantom}
The numerical validation of the ray tracing algorithms was performed using a refractive index field for which ray paths across it are known analytically. A well-known refractive index function for this purpose is Maxwell's `fish-eye lens', which is defined as
\begin{align} 
\label{eq:fish-eye}
n(\bx) = \frac{n_o}{1+(\frac{\bx}{a})^2},
\end{align}
where $n_o$ denotes the refractive index at the origin of the Cartesian coordinates ($\bx_o= [0,0,0]^T$). (Here, we set $n_o=1$ and $a=1$.) \cite{Johnson,Anderson2,Oliveira}. 
This phantom has two interesting and useful properties which will be used below to test the ray tracing algorithms' accuracy.
A ray starting from a point $p_1$ satisfying $|\bx_{p_1}| = a$ will travel along a circular path including the mirror point $p_2$ with respect to $o$, i.e., $\bx_{p_2}= -\bx_{p_1}$, and will return to the initial point on completion of the circle. Then it is clear that:
\begin{enumerate}
    \item the particular circular path followed is tangent to the initial ray direction, and 
    \item the acoustic length along the ray on completion of the corresponding circle will equal twice the acoustic length along the line segment $p_1 o p_2$.
\end{enumerate}

\begin{figure}  \centering
	\includegraphics[width=0.4\textwidth]{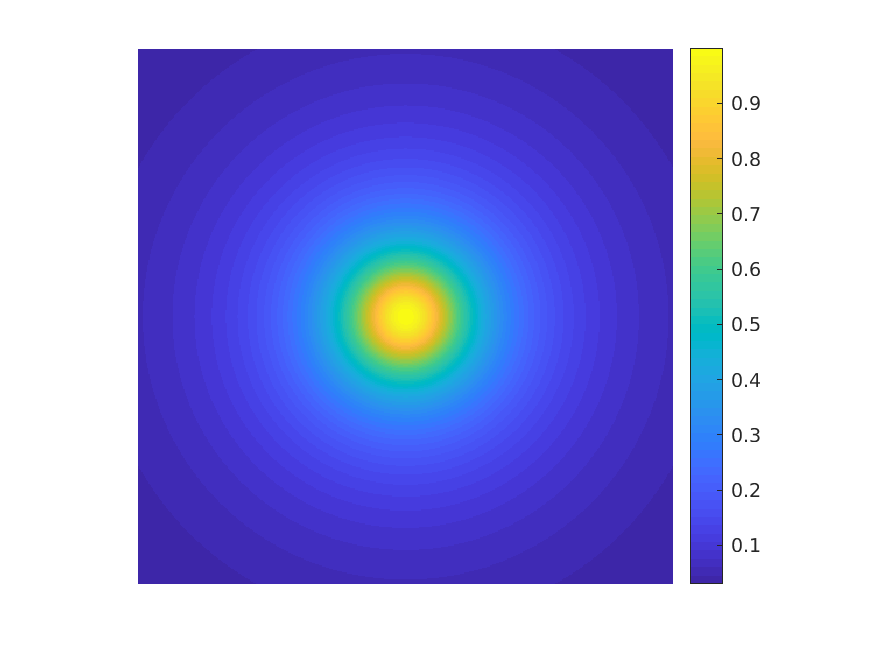}
	\caption{\textit{Maxwell's fish-eye lens}, used as a phantom for the numerical validation of the ray tracing algorithms. The phantom, which has rotational symmetry, is shown for slice $z=0$.}
    \label{fig:fish-eye}
\end{figure}

\subsubsection{Ray-path convergence}
\label{subsubsec:Ray-path_convergence}
In this section, the first property of rays propagating through the fish-eye phantom will be exploited. A family of 100 rays were initialised at position $\bx_{p_1}=[0,0,a]^T$ with initial directions tangent to a sphere centred at $\bx_{o'}=[a,a,0]^T$. The initial directions of the rays were chosen to lie in a plane normal to a line segment $p_1 o'$, and were spaced evenly to form a complete Rodriguez rotation with respect to the radial line $p_1 o'$. (In other words, 100 rotations with angles between $0$ and $2 \pi$ were applied to a reference initial direction $(-1/\sqrt{3}) [1/\sqrt{2},1/\sqrt{2},\sqrt{2}]^T$, for which the axis of rotation was the line segment $p_1 o'$.) Analytically, all the rays will travel along the surface of the sphere and finally intercept the point $p_1$. The extent to which the paths calculated using the ray tracing algorithms differ from these paths can therefore be used as a metric to test the accuracy of the algorithms.
The rays with the above initialisations were propagated using Algorithms \ref{alg:ray_tracing_dual_update} and \ref{alg:ray_tracing_mixed_step} and terminated when they reached a neighborhood of $p_1$ with a radius $\Delta s$, where $\Delta s$ is the step size. The mean radial deviation of the calculated ray points from the sphere with radius $p_1 o'$ is then computed:
\begin{align}
RE_{rd} (p_1) =  \text{mean} \left(  \frac{1}{M_{(p_1,\tilde{p}_1)} } \sum_{m=1}^{M_{(p_1,\tilde{p}_1)}}
\frac{  \text{abs}  \left( \mid \bx_{(s_m;p_1,\tilde{p}_1)} -\bx_{o'} \mid  -  \sqrt{3} a \right)  }{   \sqrt{3} a  }  \right)  \times 100.
\end{align}
Note that $\mid \bx_{p_1} - \bx_{o'} \mid = \sqrt{3} a $ and the mean operator indicates averaging across the family of 100 rays.

First, the rays were traced using analytical $n$ and $\nabla n$ (calculated from Eq.\ \eqref{eq:fish-eye}) in order to remove grid-to-ray interpolation effects associated with discretisation. The rays were traced for step sizes $\Delta s = 2^{(-4.5:0.5:3)}$ times the reference length (in Matlab notation). The reference length was chosen to be $2 \times \pi a /360$, which is equivalent to $1$ degree arc length along a circle centred at $o$.
Figure \ref{fig:deviation_convergence} shows the convergence of the error $RE_{rd}$ as the step size $\Delta s$ is decreased. It is clear that the dual-update ray tracing algorithm, Algorithm \ref{alg:ray_tracing_dual_update}, exhibits a quadratic convergence rate for $RE_{rd}$, and the traditional mixed-step algorithm, Algorithm \ref{alg:ray_tracing_mixed_step}, provides a linear convergence rate, as expected given the accuracy of the respective estimates of the direction vector updates (Sec.\ \ref{subsec:Numerical_ray_tracing}).

Second, to demonstrate the effect of the grid-to-ray interpolation on the convergence, ie.\ the effects using a discretised $n$ and $\nabla n$, the same experiment was repeated on a computational grid with spacing equal to the reference length. Figure \ref{fig:deviation_convergence} shows the convergence of the error $RE_{rd}$ in this case as the step size $\Delta s$ is decreased. As expected, for larger step sizes the discretisation effects are negligible and the algorithms provide the same convergence rate as using a continuous $n$. However, the discretisation errors become more important as $\Delta s$ decreases, eventually clearly dominating the error for Algorithm \ref{alg:ray_tracing_dual_update} for small step sizes. 

\subsubsection{Acoustic-length convergence}
\label{subsubsec:Acoustic-length_convergence}
In this section, the second property of the fish-eye phantom will be exploited. A family of 100 rays were initialised at point $p_1$ with initial directions set as described above. Therefore, in  the ideal case, all these rays would travel along the surface of the sphere, pass through the mirror point $p_2$, and finally intercept point $p_1$. Numerically, when the rays were propagated using Algorithms \ref{alg:ray_tracing_dual_update} and \ref{alg:ray_tracing_mixed_step}, they passed close to $p_2$ and then back to close to $p_1$. They were terminated when they reached a neighborhood of $p_1$ with a radius $\Delta s$, where $\Delta s$ is the step size. This final point was then discarded (as it may be beyond $p_1$) and the path was completed by linking the previous ray point to $p_1$ using a straight line. The mean error in the acoustic length was calculated using
\begin{align}  
RE_{al}(p_1) = \text{mean} \left(  \frac{  \text{abs} \big(  \bar{L}_{p_1}- L_{\text{true}}  \big)  }  { L_{\text{true}} }  \right) \times 100,
\end{align}
where the acoustic length along the rays $\bar{L}_{p_1}$ was calculated using Eq.\ \eqref{eq:perturbation_in_acoustic_length_map}, and also $\text{L}_{\text{true}}$, which is twice the acoustic length between $p_1$ and $p_2$ calculated analytically.

\begin{figure}  
    \centering
	\subfigure[]{\includegraphics[width=0.40\textwidth]{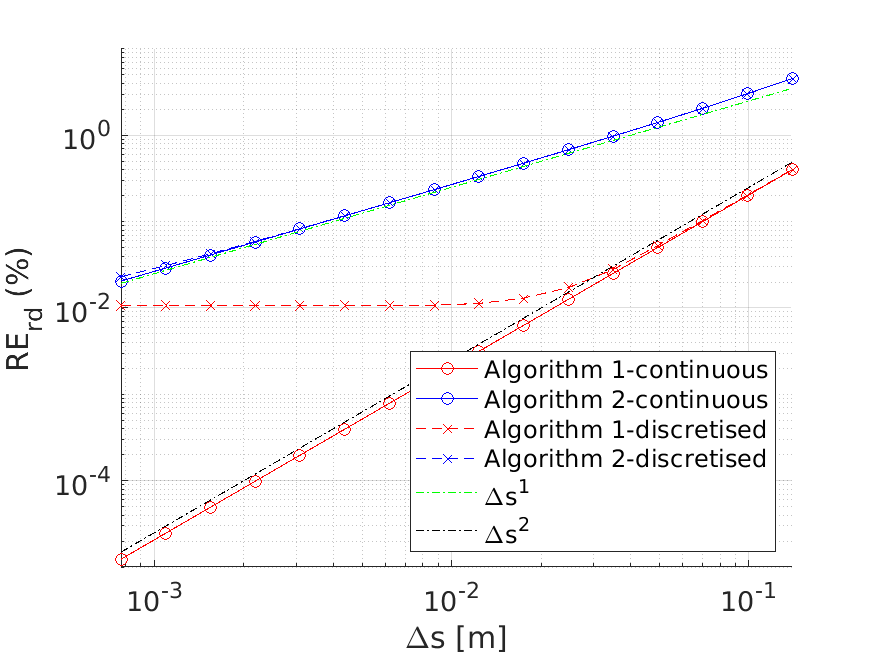}
	\label{fig:deviation_convergence}}
    \subfigure[]{\includegraphics[width=0.40\textwidth]{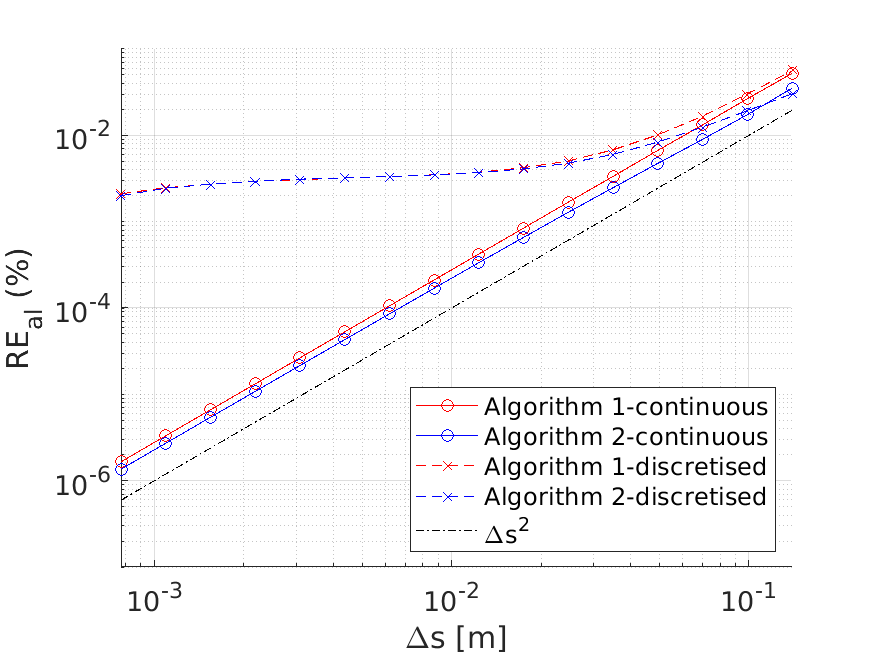}
    \label{fig:acoustic_length_convergence}}
     \subfigure[]{\includegraphics[width=0.40\textwidth]{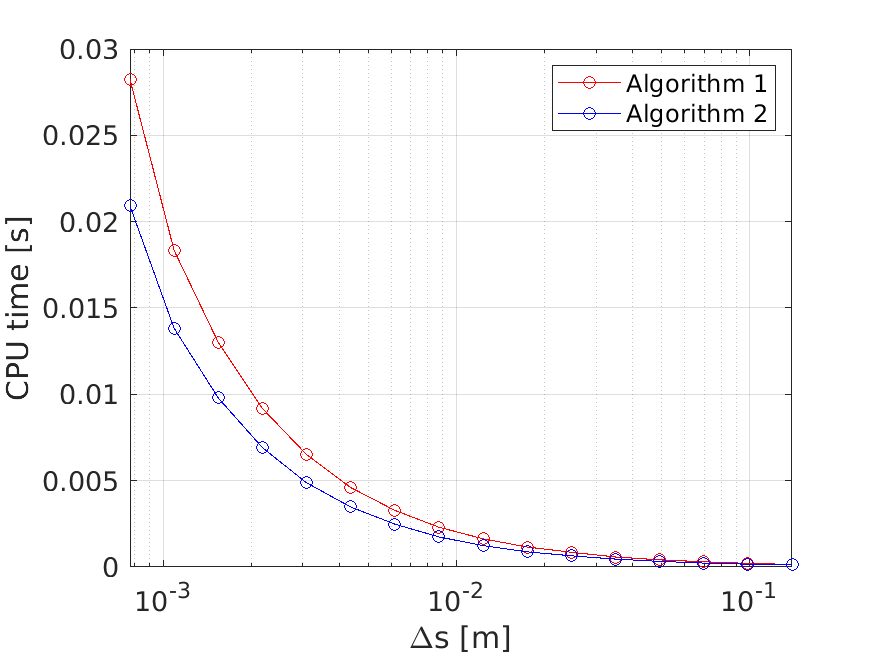}
    \label{fig:time_rays}}
	 \caption{Comparison of ray-tracing algorithms \ref{alg:ray_tracing_dual_update} and \ref{alg:ray_tracing_mixed_step}. (a) Mean error in ray-path (radius) deviation for continuous (analytical) and discretised refractive index. (b) Mean error in the acoustic length of the ray paths (integral of refractive index). (c) Mean CPU time per ray using continuous (analytical) $n$. Algorithm 2 was used in the image reconstructions in Sec.\ \ref{subsec:UST_image_reconstruction} because it is faster for the same error in the acoustic length.}
\end{figure}

Figure \ref{fig:acoustic_length_convergence} shows the error $RE_{al}$ as the step size $\Delta s$ is decreased for analytical and discretised $n$. Both ray tracing algorithms exhibit a quadratic convergence rate for $RE_{al}$, dominated by the error in the numerical integration Eq.\ \eqref{eq:perturbation_in_acoustic_length_map}. Figure \ref{fig:acoustic_length_convergence} also shows that for larger $\Delta s$ the grid-to-ray interpolation effects due to a discretisation of $n$ is negligible, but the interpolation errors dominate for small $\Delta s$, and lead to a saturation in the reduction of $RE_{al}$ with a decrease in $\Delta s$.

\newpage
\subsection{Ultrasound tomography: Data simulation}
\label{subsec:Simulation_UST_data}
\subsubsection{Imaging system}
\begin{figure}  
    \centering
	\subfigure[]{\includegraphics[width=0.45\textwidth]{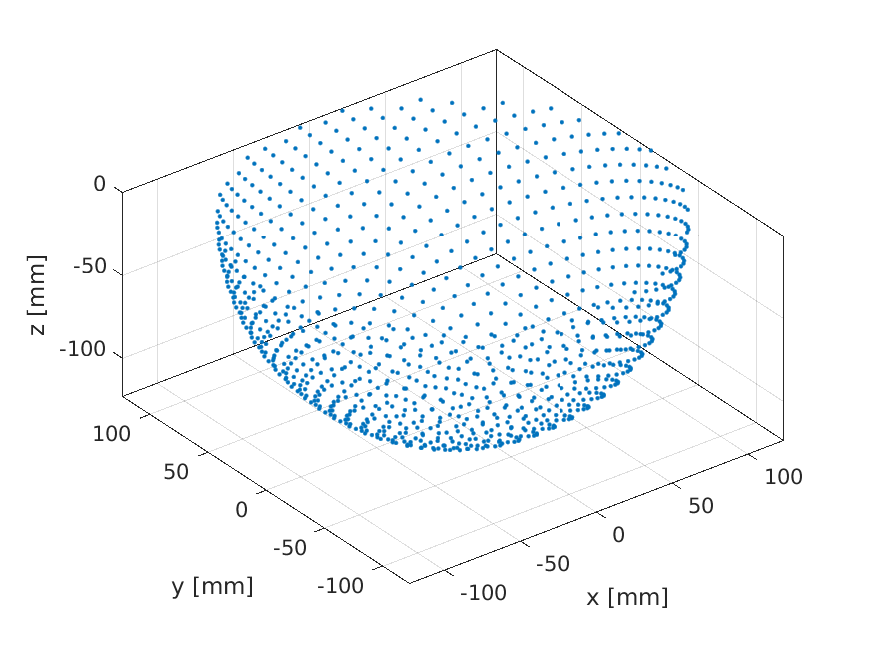}
    \label{fig:1024_emitters}}
    \subfigure[]{\includegraphics[width=0.45\textwidth]{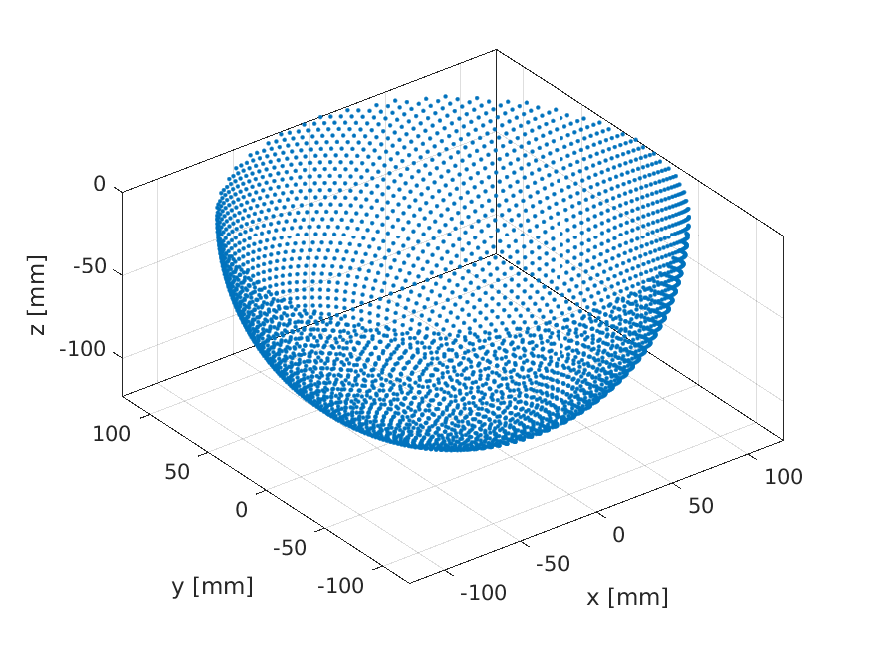}
    \label{fig:4048_emitters}}
	\caption{The hemispherical imager bowl with (a) 1024 emitters (b) 4048 receivers.}
\end{figure}

Ultrasound tomography data was simulated for an imaging system consists of 1024 emitters and 4048 receivers uniformly distributed over a hemispherical surface (bowl) of radius $R=12.35$ cm using a Golden section method. The emitters and receivers were simulated as points as shown in Figures \ref{fig:1024_emitters} and \ref{fig:4048_emitters}, respectively. 
(Some practical systems use a coarser distribution of transducers in combination with translations and rotations of the bowl to provide an equivalent amount of data \cite{Ruiter2}, as mentioned in the Introduction. Because our proposed image reconstruction approach doesn't consider the correlation between neighboring transducers, it would still be applicable when the transducers are more coarsely spaced. Note that, unlike with fixed emitters, in the rotating setting the spacing between emitters can be varied by changing the rotation angle.

\subsubsection{Digital breast phantom}
\label{subsubsec:Digital_breast_phantom}
A 3D digital phantom  mimicking the anatomical properties of the breast \cite{Lou} was used in this study. The sound speed was set to a range between $1470$ m/s and $1580$ m/s. The computational grid consisted of $522 \times 522 \times 272$ grid points ($26.1 \times 26.1 \times 13.6 $ cm$^3$ and a grid spacing of $5 \times 10^{-2}$ cm along all the Cartesian coordinates. With this grid spacing and sound speed distribution, the maximum frequency supported by the grid is $1.47$ MHz. Several slices through the phantom are shown in the left columns in Figs.\ \ref{fig:reconstructed_images_1} and \ref{fig:reconstructed_images_2}.

\subsubsection{Simulating time series data and time-of-flight picking}
A k-space pseudospectral method (k-Wave) was used for simulation of the acoustic pressure time series data \cite{Treeby}. The emission and reception points were placed on the computational grid using nearest- neighbor interpolation. Two sets of simulations were performed, the first with the bowl filled with just water (sound speed $1500$ m/s), and the second with the breast phantom in the water. To simulate the data, each emitter was individually driven by an excitation pulse, and the set of acoustic pressure time series induced at the receivers were recorded simultaneously. This was repeated for each emitter. The pressure time series were recorded at 4245 time points with a sampling rate of $20$ MHz, and additive white Gaussian noise was added to give a $40$ dB signal-to-noise ratio. The computational time for simulation of the two sets of data using k-Wave's GPU code \cite{kWave} on eight NVIDIA Tesla P40 Pascal GPUs was about 7.5 days.

\begin{figure}     
    \centering
    \subfigure[]{\includegraphics[width=.49\textwidth]{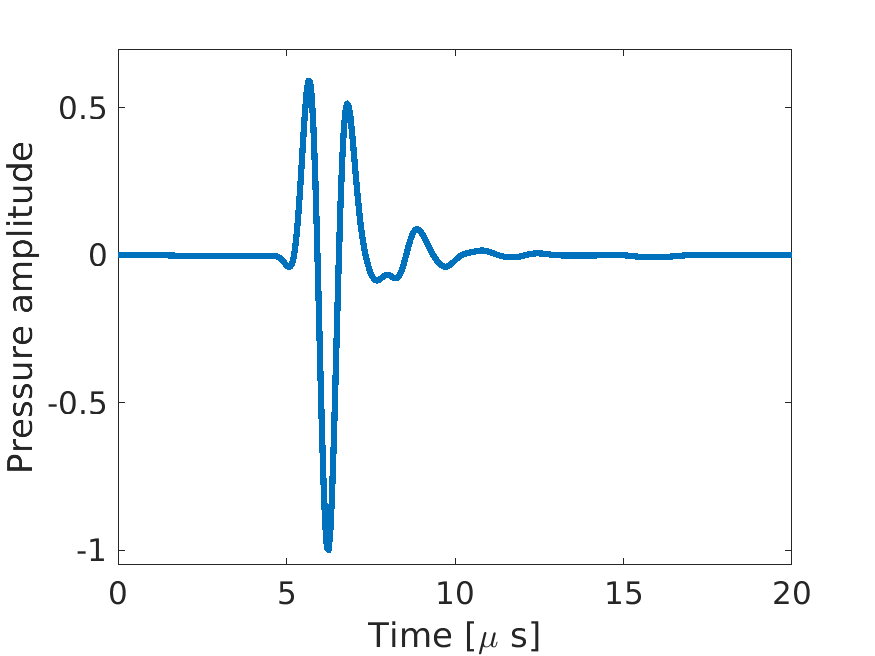}\label{fig:pulse_time}}
    \subfigure[]{\includegraphics[width=.49\textwidth]{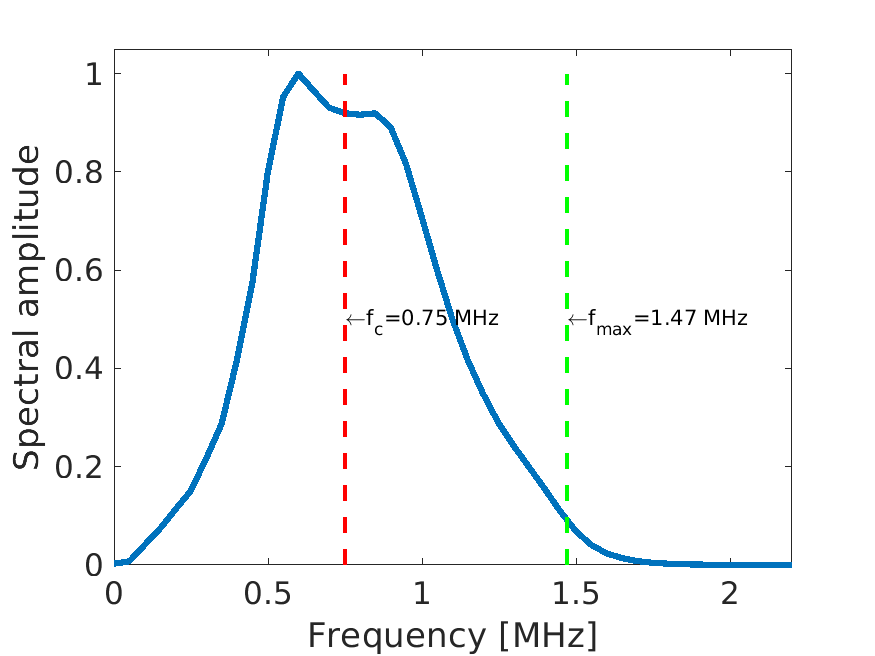}\label{fig:pulse_spectrum}}
    \caption{(a) The excitation pulse, and (b) its frequency spectrum. The red dash line shows the centre frequency, and the green dash line shows the maximum frequency supported by the computational grid.}
\end{figure}

The output of practical ultrasound transducers covers a finite (and typically quite limited) bandwidth, and the field produced becomes more directional at high frequencies. In choosing transducers for a imager, therefore, a trade-off must be made between the range of frequencies in the excitation pulse and the directionality of the detectors. In practical breast imaging systems, because acoustic absorption attenuates the higher frequencies preferentially, lower frequencies tend to be preferred, (typically below 5 MHz). The system modelled here uses relatively low frequencies ($0.75$ MHz centre frequency) and assumes the emitters and detectors are omni-directional. The excitation pulse and its spectrum are shown in Figs.\ \ref{fig:pulse_time} and \ref{fig:pulse_spectrum}.

There are three principal, practical, reasons for choosing this frequency range for the simulations. First, to match an experimental imager that is currently under construction. Second, to limit the time taken to simulate the 3D data to about one week
given the available hardware. Third, when time-of-flight images, such as we are producing here, are used as a starting point for full-wave inversions, they must be able to use the time series recorded for the full-wave inversions. When performing 3D image reconstruction with full-wave methods it is currently impractical to use higher frequencies over volumes of this size as the forward and adjoint calculations are too computationally demanding. Lower frequencies, however, can be modelled.

It is immediately clear that the effect of the chosen frequency range is that the pulse is not tightly confined in time. When it has propagated through the heterogeneous breast medium from the emitter to the detector, the long pulsewidth makes it harder to pick the time-of-arrival of the pulse at the detector. This is a non-trivial challenge in practice, and there are many methods devoted to accurate time-of-flight picking. The method used here, and described in the Appendix, was found to be robust for this simulated breast imaging time series data.

It is important to note that the performance of ray-based approaches relying on first-arrival picking depends strongly on the properties of the excitation pulse. The decision to use lower frequency excitation pulses therefore has the effect of introducing more uncertainty into the time-of-flight data that is used as the input to the image reconstruction. When it is possible to use excitation pulses with higher frequencies and wider bandwidths, the time-of-flight picking would be expected to be more precise, having a knock-on beneficial effect on the reconstructed images.

\subsection{Ultrasound tomography: Image reconstruction procedure}
\label{subsec:UST_image_reconstruction}
The grid used for reconstruction consisted of $262 \times 262 \times 138$ grid points with a spatial spacing of $1$ mm along all the Cartesian coordinates. (This is different from the grid used to simulate the data to avoid the \textit{inverse crime}, although this study doesn't use the same model for data generation and image reconstruction anyway.)

Prior to the image reconstructions, a Region of Interest (ROI) that encompassed the heterogeneous breast region was found, and the refractive index outside this region was set to $1$. To define the ROI, a rough image was computed using straight rays and a binary image segmentation scheme based on an edge-based active contour method \cite{Caselles} was applied on the reconstructed low-contrast image to separate the breast region (ROI) from non-breast. 
The initial guess for the refractive index was set to $\bar{n}=1$, which corresponds to the known sound speed in water ($1500$ m/s). For each emitter, rays are computed only for receivers that are at least $8$ cm from the emitter, based on the idea that a ray linking closer points will not hit the breast region, and is thus not useful in the image reconstruction. The image reconstructions were performed in Matlab R2019a (The MathWorks, Inc) using one 8-core Xeon E5-2620 v4 2.1 GHz CPU. 

\subsubsection{Straight-ray approach}
For comparison with the bent-ray approach, the breast phantom was reconstructed using straight rays, i.e.\ assuming that the medium is non-refracting and so the rays travel along straight lines between the emission and reception points. To implement this, the initial direction $\boldsymbol{d}_{\bar{e}}$ in Algorithm \ref{alg:ray_tracing_mixed_step} was set to the unit vector pointing from the emission point to the reception point, and the ray's path was calculated using $\nabla n = 0$ and $\Delta s= 1$ mm, the grid spacing for image reconstruction. Because for this case, the forward operator $A$ is linear, the refractive index was found by minimising the least squares problem  Eq.\ \eqref{eq:nonlinear_minimisation_for_n} using a steepest descent algorithm, as explained in Section \ref{subsec:Inverse_problem}. The sequence of reconstructed images were calculated using a constraint for $\bar{n}$ outside the ROI, and the difference from the true refractive index was recorded after every 40 steepest descent iterations for comparison with the bent-ray approaches. The straight-ray inversion algorithm was terminated when the stopping condition in Eq.\ \eqref{eq:stopping_criterion} was satisfied. 

\subsubsection{Bent-ray approaches}
\label{subsubsec:bent_ray_approaches}
In the bent-ray approaches, the UST inversion for the refractive index was solved as a sequence of linearised minimisation problems, Eq.\ \eqref{eq:qth_linearised_minimisation}, as described in Sec.\ \ref{subsec:Inverse_problem}. For each linearised subproblem $q$, the Jacobian matrix $J$ was formed as described in Sec.\ \ref{subsubsec:Jacobian_matrix} using the ray paths between all sufficiently separated emitter-receiver pairs. Each linear subproblem was terminated after 400 steepest descent iterations. 

Algorithm \ref{alg:ray_tracing_mixed_step} was used with $\Delta s= 1$ mm to trace rays between each pair of emitters and receivers. This was implemented with $\nabla n$ zero outside the ROI, assuming the medium outside the ROI is homogeneous, to reduce the computational cost. For the first iteration in the UST problem, iteration $q=0$, the refractive index was set to be homogeneous ($\bar{n}=1$) and thus the first traced ray reaches the corresponding reception point directly (as in the straight-ray case). For subsequent iterations, the discretised refractive index was smoothed by convolving with a cube with sidelength equalling three grid points in all Cartesian directions.

For each ray, a ray-linking algorithm was then used to match the end point of the rays to the corresponding reception points. Images were reconstructed for all three ray-linking approaches. 
The ray-linking performances of the three ray-linking approaches are compared in Sec.\ \ref{subsec:comparison_ray-linking_algorithms} below, and the final reconstructed images are given in Sec.\ \ref{subsec:Ultrasound_tomography_Results}.

\subsection{Comparison of ray-linking algorithms}
\label{subsec:comparison_ray-linking_algorithms}

The image reconstructions described above were performed for all three ray-linking algorithms, and the accuracy and convergence rates with which the algorithms honed in on the target reception points were compared. All the rays were traced as described in Sec.\ \ref{subsubsec:bent_ray_approaches} above, and, for each ray, one of the ray-linking algorithms was used to match the end point of the ray to its corresponding reception point. All the ray-linking algorithms were terminated when the residual function $\mathcal{E} \big[ \bd_{\bar{e}}^k \big] $ reached a tolerance $\varepsilon_\text{link} = 1 \times 10^{-6}$, or if the number of ray-linking iterations $k$ exceeded $N_\text{link}=100$. 

\subsubsection{Gauss-Newton}
Ray-linking with the Gauss-Newton approach was performed using Algorithm \ref{alg:ray_linking_Gauss-Newton}. For each iteration $k$, Algorithm \ref{alg:adaptive_smoothing_Gauss_Newton} was used to adjust the perturbations enforced on the initial angles for calculation of the finite differences, Eq.\ \eqref{eq:Jacobian_finite_difference}, in the Jacobian matrix $\mathcal{J}$ via controlling the parameter $\tau$. The algorithm was initialised by $\tau=1 \times 10^{-15}$ for all $k$, and was recursively increased by a factor $\eta=10$ until the conditions in Eq.\ \eqref{eq:singular_value_conditions} were satisfied, with $\vartheta = 1 \times 10^4$ and $\varsigma = 1 \times 10^{-4}$. This loop was unconditionally terminated when the associated iteration number $k_h$ became greater than $N_h=5$, i.e., when an increase of $\tau$ up to $10^{-10}$ still does not sufficiently reduce the ill-conditioning of $\mathcal{H}^k$. In these rare cases, the ray-linking problem was restarted using an initial $\tau$ for all $k$ 10 times greater than the previous attempt. (For evaluation purposes, the iteration number $k$ was not reset.) After calculation of a well-conditioned Hessian matrix $\mathcal{H}^k$, the initial direction was updated using Eq.\ \eqref{eq:d_line_search} with a step length $\alpha^k$ satisfying Eq.\ \eqref{eq:Armijo_condition} with $\mu = 1\times 10^{-4}$. 

\subsubsection{Quasi-Newton} 
The Quasi-Newton ray-linking approaches were performed using Algorithm \ref{alg:ray_linking_Quasi_Newton}. The matrix $\mathcal{B}^0$ was calculated using Eq.\ \eqref{eq:Jacobian_finite_difference} with a fixed perturbation of $1 \times 10^{-6}$ for both components. The update directions were bounded using upper and lower bounds $\boldsymbol{l}= \bd_{(0,q)}^0 -0.2 $ radians and $\boldsymbol{u}= \bd_{(0,q)}^0 + 0.2 $ radians for all $q$, and $\zeta=0.5$,  $\kappa = 1 \times 10^{-6}$. 
For the \textit{BFGS-like} approach, the approximate derivative matrix $\mathcal{B}^k$ was updated using Algorithm \ref{alg:adaptive_smoothing_BFGS} with the parameter $\tau$ initialised by 1 and recursively reduced by a factor $\eta =0.5$ until $\mathcal{B}^k$ satisfied the conditions in Eq.\ \eqref{eq:singular_value_conditions}. (The conditions used the same parameters as for the Gauss-Newton method except that the objective function value $\mathcal{E}\big[\bd_{\bar{e}}^k\big]$ was replaced by its gradient). The smoothing loop was unconditionally terminated if the iteration number $k_h$ exceeded $N_h= 20$. 
For the \textit{Broyden-like} approach, the derivative matrix $\mathcal{B}^k$ was updated using Algorithm \ref{alg:adaptive_smoothing_Broyden}. The parameter $\tau^k$ was initialised as 1 and changed within a neighborhood $\bar{\tau}=0.1$ with an increment parameter $\eta_c=\pm 0.01$  until the conditions in Eq.\ \eqref{eq:singular_value_conditions} were satisfied with the same parameters as used for the BFGS-like approach.

\subsubsection{Numerical results} 
\begin{figure}    
    \centering
    {
    \subfigure[]{\includegraphics[width=.45\textwidth]{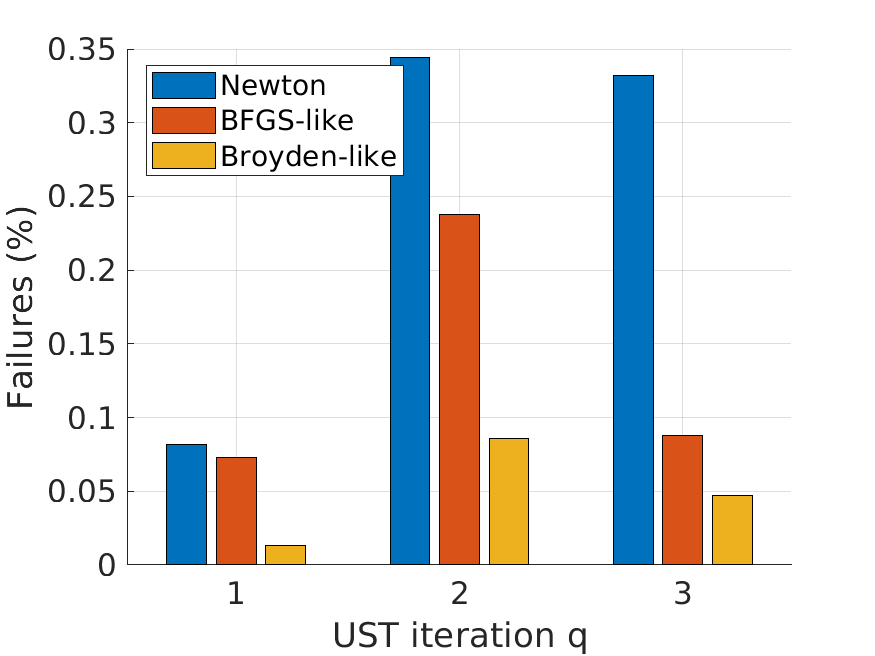}
    \label{fig:ray_linking_failures}}
    \subfigure[]{\includegraphics[width=.45\textwidth]{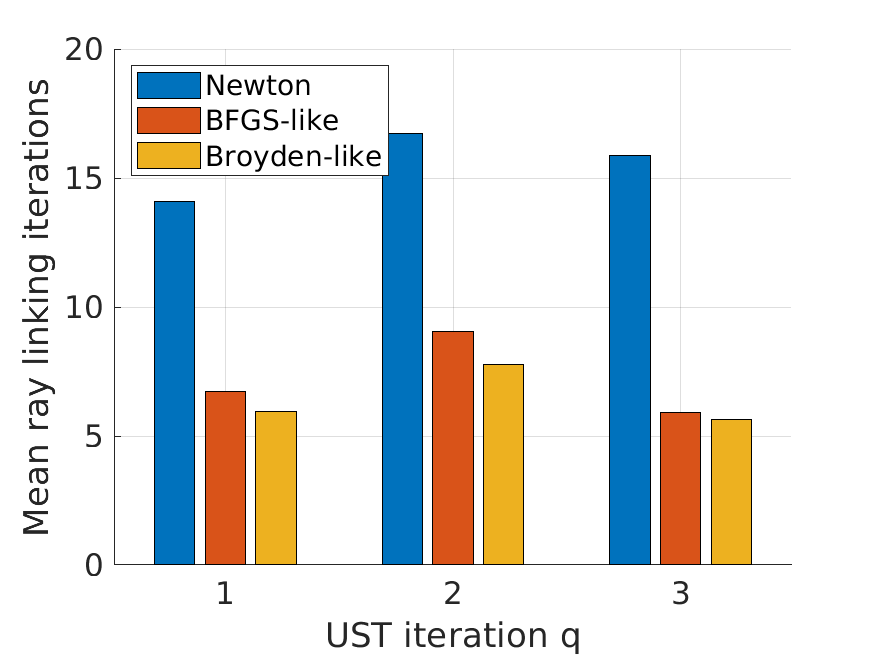}
    \label{fig:ray_linking_iterations}}
    }
    \caption{Ray-linking comparisons: (a) failure rate (\% of refracted rays that failed to reach to within the required tolerance of the reception point within the maximum number of iterations), (b) mean number of ray-linking iterations required.} 
    \label{fig:ray_linking_comparison} 
\end{figure}

Figure \ref{fig:ray_linking_failures} shows the fraction of the \textit{refracted} rays for which the end point failed to reach sufficiently close to the reception point within the maximum permissible number of iterations. The failures were calculated as a percentage of the number of rays that hit the breast phantom and were therefore refracted, not as a percentage of all emitted rays. (If the non-refracted rays passing through the water only were included, this percentage would be much lower.) The results are shown for iterations $q=1,2,3$. Subproblem $q=0$, in which all the rays are straight rays, does not appear in Figure \ref{fig:ray_linking_failures}, and the stopping criterion of our UST inverse problem, Eq.\ \eqref{eq:stopping_criterion}, was satisfied at iteration $q=3$.

At the start of each iteration $q$, the rays need to be initialised. Rays previously successfully linked (on iteration $q-1$) used the same initial direction for iteration $q$. For rays that failed to link previously, a decision was made between using the last initial direction reached on iteration $q-1$, 
$\bd_{(\bar{e},q-1)}^{N_{\text{link}}}$,
and the first one, $\bd_{(\bar{e},q-1)}^0$, the ray 
giving the smaller value of the functional
$\mathcal{E}$, Eq.\ \eqref{eq:ray_linking_minimisation}, being used.
The reason for not discarding outright the ray that failed to link is that, typically, the algorithm has not diverged (because the singular values are controlled) but is stuck in a local minimum often, but not always, close to the reception point.

As shown in Figure \ref{fig:ray_linking_failures} for all ray-linking algorithms, the fraction of rays that failed to reach the termination tolerance was very low at less than $0.5  \%$. Among these algorithms, the Broyden-like approach performed best with fraction of failures on the order of $10^{-4}$ for all UST iterations. 

Note that it would be possible to re-attempt the failed ray-linking problems after changing the size of the smoothing window enforced on the refractive index distribution (just for those specific rays). Because the portion of failures is very low, the increase in the computational cost of the whole image reconstruction algorithm would be negligible. Nevertheless, this approach was not taken in this work.

Figure \ref{fig:ray_linking_iterations} shows the mean number of ray-linking iterations taken for a ray to reach sufficiently close to the reception point.
This includes the rays that failed to reach the stopping threshold after $N_\text{link}$ iterations, but not the unrefracted rays that reach the reception points directly, as these rays do not require ray-linking. For the Gauss-Newton method, each iteration required at least 3 rays to be traced, 2 for the Jacobian and one for the residual, and maybe more in rare cases when smoothing loop and the backtracking loop require additional function evaluations.
For the Quasi-Newton methods, in contrast, each iteration was equivalent to tracing only 1 single ray. The mean number of iterations for ray-linking using the \textit{Broyden-like} approach was about 6.

\subsection{Ultrasound tomography: Results}
\label{subsec:Ultrasound_tomography_Results}

\subsubsection{Reconstructed sound speed (refractive index) images}

\begin{figure}     
    \centering
    {
    \includegraphics[width=.49\textwidth]{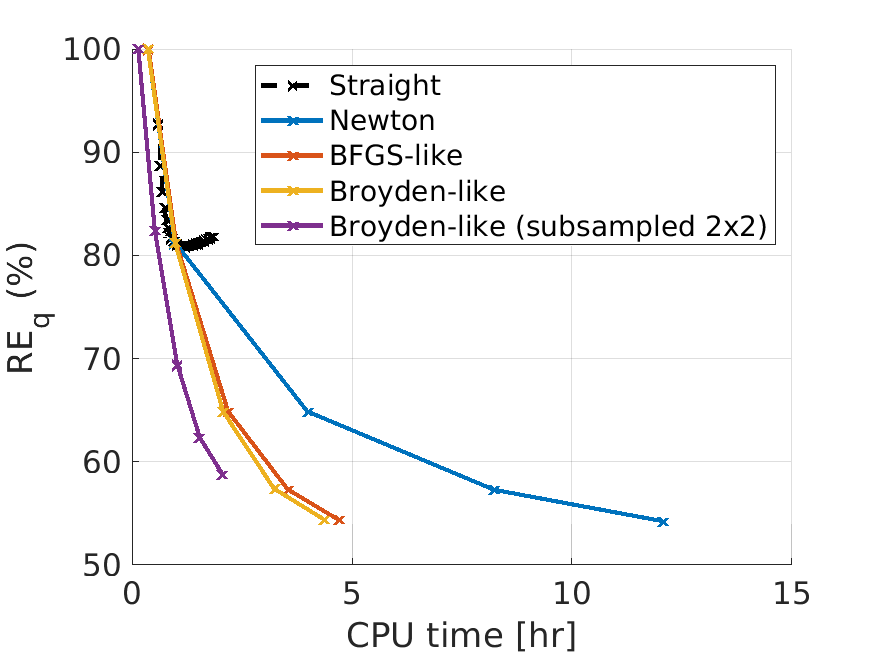}
    }
    \caption{Accuracy of the reconstructed images against the computation time (hours) for the various reconstruction approaches, the relative error at each iteration $RE_q$ 
    } 
    \label{fig:error_vs_CPU_time} 
\end{figure}
\begin{figure}     
    \centering
    \includegraphics[width=\textwidth]{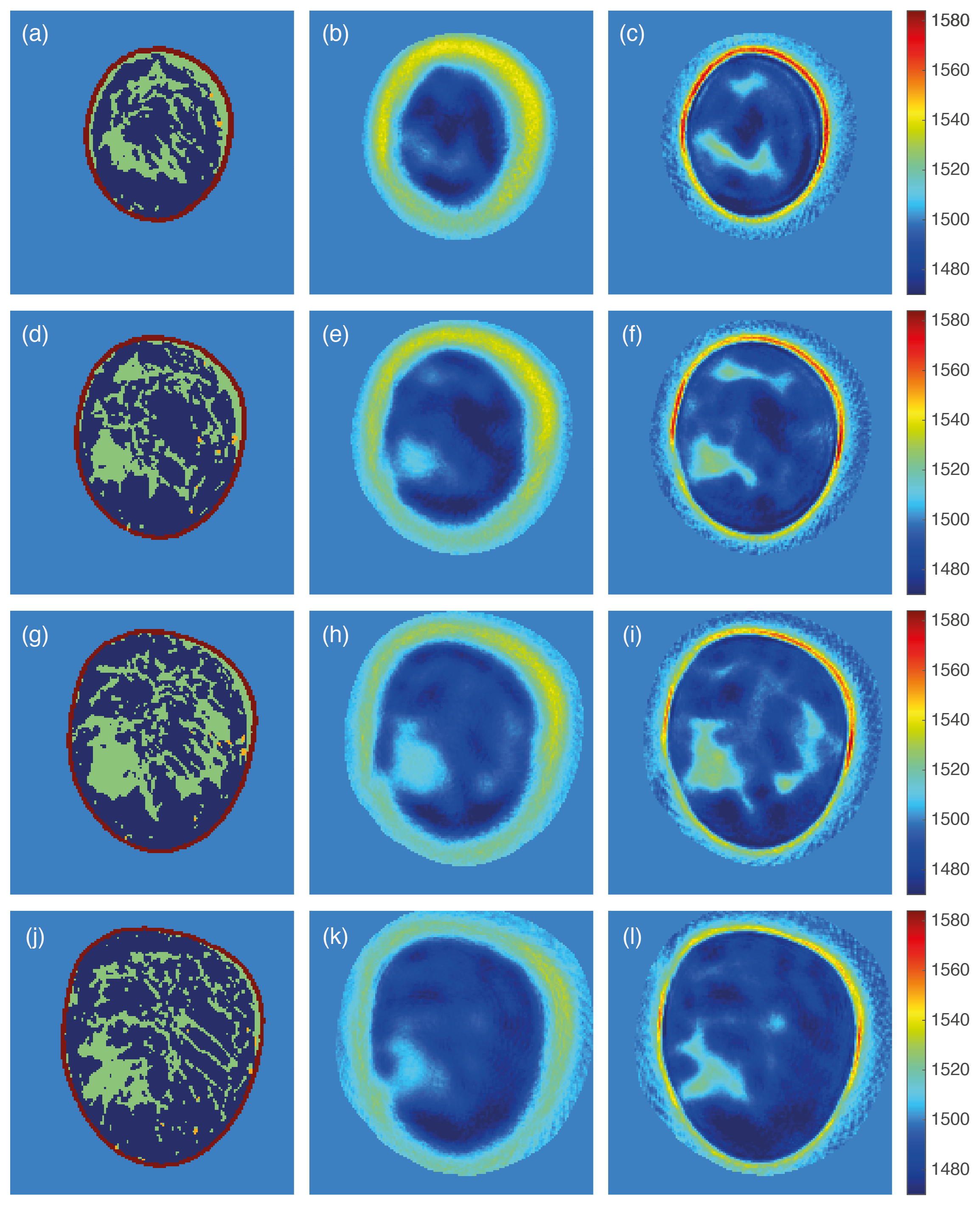}
    \caption{Left: true sound speed distribution, Centre: reconstructed using straight-ray approach, Right: reconstructed using bent-ray approach. Slices through the 3D images at (a)-(c): $z=-7$ cm, (d)-(f): $z=-6$ cm, (g)-(i) $z=-5$ cm, (j)-(l) $z=-4$ cm.} 
    \label{fig:reconstructed_images_1}
    \end{figure}
\begin{figure}     
    \centering
    \includegraphics{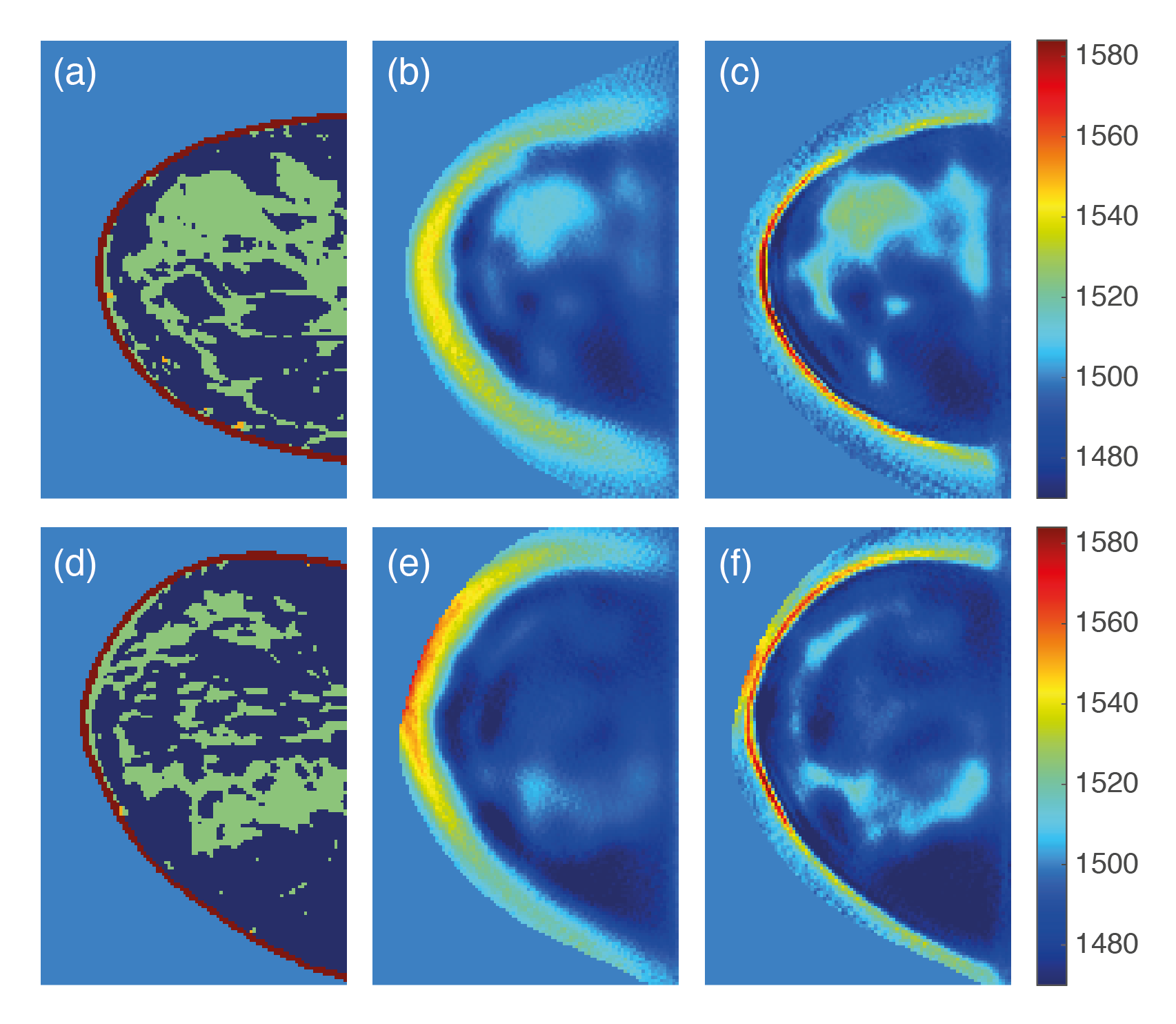}    
    \caption{Left: true sound speed distribution, Centre: reconstructed using straight-ray approach, Right: reconstructed using bent-ray approach. Slices through the 3D images at (a)-(c): $x=0$ cm, (d)-(f): $y=-1$ cm.} 
    \label{fig:reconstructed_images_2}
\end{figure}

Figures \ref{fig:reconstructed_images_1} and \ref{fig:reconstructed_images_2} show slices through the 3D image volumes of the reconstructed breast phantom. 
The left column in Fig.\ \ref{fig:reconstructed_images_1}
shows slices through the phantom at depths of $z = {-7, -6, -5,-4}$ cm; the left column in Fig.\ \ref{fig:reconstructed_images_2}
shows slices through the phantom at $x = 0$ and $y = -1$ cm.
The centre columns show the images reconstructed using straight rays (the images for which $RE_q$ was minimum are shown), and the right columns show the images reconstructed using the bent-ray code. (Only the images reconstructed using the fastest ray-linking approach, \textit{Broyden-like}, are shown as all three approaches gave very similar images). All the figures are shown on the same colour scale. A metric was used to quantifying the accuracy of the reconstructed images, the squared relative error,
\begin{align}
RE_q= \frac{\| \bar{c}_{q}- \bar{c}_{breast} \|_2^2}{ \|  \bar{c}_w - \bar{c}_{\text{breast}} \|_2^2} \times 100 \%,
\end{align}
where $\bar{c}_{q}$ is the reconstructed images of the sound speed after the subproblem $q$ and $\bar{c}_{\text{breast}}$ is the sound speed of the breast phantom interpolated to the image reconstruction grid. 
Figure \ref{fig:error_vs_CPU_time} shows $RE_q$ against the elapsed computation time for the whole image for each approach tested: straight rays and the three bent-ray methods (Gauss-Newton and the two Quasi-Newton methods). The Broyden-like Quasi-Newton method was also used using data evenly subsampled by a factor of $2$ for both emitters and receivers (a four-fold reduction in emitter-receiver pairs). The computation time in this plot includes the calculation of the ROI, the construction of the Jacobian matrices, and the steepest descent iterations for all subproblems $q$, but not the time-of-flight picking. Because an accelerated data processing approach was used (see Appendix) the time for calculation of the times-of-flight was just a few minutes using two CPUs. 
The results in Figs.\ \ref{fig:reconstructed_images_1} and \ref{fig:reconstructed_images_2} show the clear improvement that the bent-ray approach has over the straight ray approach, both in terms of image resolution and quantitative accuracy. Furthermore, as Fig.\ \ref{fig:error_vs_CPU_time} shows, the sub-sampled bent-ray approach has reduced the squared error, compared to the straight ray case, by about 25\% for about the same computation time and with four times fewer transducers.

\clearpage
\section{Discussion and Conclusions} 
\label{sec:Discussion_Conclusions}
Ultrasound tomography to obtain sound speed maps of biological tissue has been receiving increasing attention in recent years, especially for breast imaging, as highlighted in the Introduction. This is due to the greater availability of high channel count ring and bowl-shaped arrays and the huge growth in available computational power \cite{Ruiter2,Hopp1}. The latter has also led to image reconstructions based on full-wave inversion methods becoming a popular topic of research. However, for 3D UST reconstructions full-wave approaches are computationally very intensive and still in their infancy. Furthermore, the inversions depend strongly on an accurate initial guess of the sound speed distribution as artefacts in the initial guess can be difficult to eradicate \cite{Wiskin2}.
There is a need, therefore, for a more robust and quicker approach to 3D UST image reconstruction that can provide a low-artefact, quantitatively accurate, sound speed map as a starting point for full-wave UST inversion methods \cite{Wiskin1,Wiskin2}.
Such a scheme would have other uses too. Photoacoustic tomography, which has similarly been receiving a great deal of attention in the past decade for 3D breast imaging \cite{Lin,Nyayapathy}, depends on quantitatively accurate knowledge of the sound speed distribution. Even a low resolution (smoothed) sound speed map, if quantitatively accurate, could significantly improve 3D photoacoustic tomography reconstructions in heterogeneous media such as breast, where the image extends to deeper than superficial regions.

3D UST using ray-based methods can fulfil the needs of both full-wave UST and photoacoustic tomography. Ray-based UST methods are computationally more efficient than full-wave methods, and they are more robust, ie.\ much less prone to model-mismatch in, for example, the source pulse shape or the exact characteristics of the detectors. Also, 3D UST image reconstructions based on rays can, of course, be used as the primary image reconstruction approach (if the resources for implementing full-wave UST are not available, for instance).

\subsection{Contributions of this work}
This paper presents a comprehensive derivation and study of a robust framework for large-scale bent-ray UST in 3D for a hemispherical detector array. While ray-based methods have been studied for decades, in geophysics in particular \cite{Rawlinson2}, most previous work for biological tissue has been restricted to 2D geometries or to obtaining reflection images. In particular,
\begin{itemize}
    \item This is the first study to present an algorithm for quantitative sound speed estimation in 3D using a hemispherical bowl array, and tailored for UST of breast. 
    \item \textbf{Ray-tracing.} It was shown that a \textit{dual-update} approach (Algorithm \ref{alg:ray_tracing_dual_update}) to ray-tracing provided more accurate ray-paths than the traditional \textit{mixed-step} approach (Algorithm \ref{alg:ray_tracing_mixed_step}), but for calculation of the acoustic length, which is dominated by the step size along the ray, the algorithms performed similarly. Algorithm \ref{alg:ray_tracing_mixed_step} was more computationally efficient. 

    \item \textbf{Ray-linking.} The ray-linking problem in 3D is challenging due to the high number of degrees of freedom for the trajectory of rays \cite{Rawlinson2}. An approach to ray-linking was proposed that is novel in the way the problem is parameterised and solved for a hemi-spherical detection surface, which could be straightforwardly generalised to any convex detection surface. This problem was then analysed both as a minimisation, Eq.\ \eqref{eq:ray_linking_minimisation}, and as a root-finding problem, Eq.\ \eqref{eq:ray_linking_root_finding}. A \textit{Gauss-Newton} approach was proposed to solve the former, and two derivative-free 
    \textit{quasi-Newton} approaches (\textit{Broyden-like} and \textit{BFGS-like}) were proposed to solve the latter. A limitation for these \textit{quasi-Newton} approaches is the lack of efficient line search techniques. Here, a simple box constraint for mitigating nonmonotone behaviour and avoiding divergence was proposed and found to work well. 
    \item \textbf{Adaptive smoothing.} To overcome ill-posedness in the ray-linking problem, novel adaptive smoothing schemes were proposed. These controlled the conditioning of the Hessian in the \textit{Gauss-Newton} scheme, and the approximate derivative matrix in the \textit{quasi-Newton} methods to ensure successful ray-linking. The fraction of ray-linking failures using the \textit{Broyden-like} approach is about $0.05\%$ (Fig.\ \ref{fig:ray_linking_comparison}). (This neglects the pairs for which the first traced ray hits the reception point directly.) The mean number of rays needed for ray-linking between an emitter-receiver pair was only about 6. 
    \item \textbf{Realistic simulation design.} The approach and algorithms presented here are intended to be practically useful, hence the emphasis is on robustness and efficiency at each stage of the algorithm design. In order to ascertain their likely performance with real measured data from a breast imager, a highly realistic scenario was simulated. First, the bowl array and number of transducers were chosen to be of comparable size and number to those used in practical breast imaging systems. Second, the phantom used was realistic of breast tissue \cite{Lou}. Third, the frequency spectrum of the excitation pulse was bandlimited, mimicking the frequency range of detected signals in practice, with a relatively low centre frequency. (This was partly for data generation reasons - see below - and partly so that the same data could later be used for 3D full-wave inversion, which is also restricted in the usable frequency range for computational reasons.)
    Fourth, the data was simulated using a broadband full-wave model \cite{Treeby} and a realistic level of noise was added to the time series. Fifth, a commonly-used TOF-picking algorithm (with a minor modification for speed-up and reduction of mispicks) was used, so the TOF values contained a realistic level of variance - see appendix. 
    \item \textbf{Image reconstruction.} The nonlinear UST problem of estimating the sound speed was solved as a series of linearised subproblems, each solved using a steepest descent scheme. (Other methods, eg.\ based on conjugate gradients or LSQR, could also be used but in our experience the images produced by these approaches are similar, and, for practical cases, they are more sensitive to the stopping criterion used for termination of each linearised subproblem.) Figure \ref{fig:error_vs_CPU_time} shows that in about the same computational time as the straight ray case,
    a sub-sampled \textit{Broyden-like} bent-ray approach reduced the squared error by about 25\% with just 25\% of the transducers.
    A comparison of the images in Figs.\ \ref{fig:reconstructed_images_1} and \ref{fig:reconstructed_images_2} shows that the bent-ray approach offers a significant improvement over the straight-ray approach.
\end{itemize}

\subsection{Limitations}

As explained in Sec.\ \ref{sec:Ray_tracing}, the theory of ray tracing is based on a high-frequency approximation. UST systems designed for use with ray-based image reconstructions therefore often use excitation pulses with a central frequency of about $2.5$ MHz \cite{Duric,Ruiter,Ruiter2}. (The limit is the preferential high frequency attenuation of tissue.) 
Here, an excitation pulse with a much lower central frequency of $0.75$ MHz was used in order to manage the computational cost of running full-wave acoustic simulations for $1024$ emitters. Even so, these 1024 3D simulations took more than a week on an 8-GPU cluster (see Section \ref{subsec:Simulation_UST_data}) with a fast full-wave simulation code \cite{kWave}.
The TOF-picking, and therefore the accuracy of the reconstructed images, would be significantly improved were a more broadband excitation pulse with a higher central frequency used (as may be the case with experimentally measured data). Unusually, therefore, for a simulation study, the images shown here are unlikely to represent the best performance of the algorithms, but are limited by the 3D data simulation.

The number of emitters was also limited by the computational demands, and so it was not practical to study a scenario with a greater number of emitters, which are the case for example in a translational-rotational setting geometry \cite{Ruiter2}, to see how it would affect image accuracy. By studying what happens when the number of emitters is reduced, it seems likely that there would be some improvement with more emitters but not a dramatic change.

There are several steps in the algorithms where approximations are made. Any of these could, under various circumstances, become the limiting factor in the image accuracy, although for the simulations shown here the variance in the TOF estimates dominates the error. These include the window used to smooth $\bar{n}$, grid spacing and grid-to-ray interpolation effects (the shape function used for trilinear interpolation is non-differentiable, but represents a trade-off between accuracy and speed), spacing along the ray, and the stopping criteria.

\subsection{Summary}
A set of robust and efficient algorithms have been designed for 3D bent-ray ultrasound tomography of the breast (ray tracing, ray-linking, and solving for the sound speed). These have been validated and tested with realistic simulated data. Because of the computational demands in generating the data for the simulation, the algorithms could not be tested to the limit of their performance. Nevertheless, the results demonstrate the usefulness of these algorithms for breast tomography and the technique could be used either as an adjunct to a full-wave inversion or another modality such as photoacoustic tomography, or as an efficient image reconstruction approach in its own right.

\section*{Acknowledgements}
This work was funded by the European Union’s Horizon 2020 Research and Innovation program H2020 ICT 2016-2017 under Grant agreement No. 732411, which is an initiative of the Photonics Public Private Partnership. The authors would like to thank Felix Lucka, Michael Jaeger, and others in the Pammoth consortium, \url{https://www.pammoth-2020.eu}, as well as Marta Betcke, Francesc Rul$\cdot$lan, Bradley Treeby and UCL's Biomedical Ultrasound Group for helpful discussions on various aspects of ray-based inversions and numerical modelling.

\appendix

\section{Time-of-flight picking algorithm}    
\label{sec:appendix}

This appendix describes the method used for picking, from the measured time series, the time of the first arrival of the ultrasound pulse after it has travelled through just water, $\bar{T}_w$, and after it has travelled through the object in water, $\bar{T}_{\text{object}}$ (see Sec.\ \ref{subsec:Inverse_problem}). In this approach, based on \cite{Li3}, the time at which the Akaike Information Criterion (AIC) is minimum is calculated in a time window that is expected to include the first arrival of the signal. This method is called \textit{Modified AIC}. Our experience with signals simulated from the breast phantom, Sec. \ref{subsubsec:Digital_breast_phantom}, and real signals measured in a 2D UST experiment, is that this approach outperforms many other first arrival picking approaches found in the geophysical literature, for example those reviewed in \cite{Akram}. We also found that the accuracy of the \textit{Modified AIC} approach depends on the chosen time window. 

Here, we propose an approach for selecting an appropriate time window by including information about the amplitude of the envelope of the measured signal. We confirmed the effectiveness of our approach on measured 2D UST data (not shown). Each measured signal is defined using $\boldsymbol{y} $ with indices $i \in \left\{1,...,N_t \right\}$, and $t_i$ denoting the measurement time instant associated with the index $i$.

\textbf{Step 1: Envelope.} Normalise the amplitude of the signal, and then calculate the envelope (absolute Hilbert transform).
    
\textbf{Step 2: Large window.} For each measured signal, based on the distance between the relevant emitter and receiver pair together with an assumption for the minimum and maximum sound speed, choose a \textit{large time window}.
Here, $1400$ and $1600$ m/s were used for the minimum and maximum values for the sound speed. (In practical cases it might be possible to reduce the difference between these values; here the extremes are used, ie.\ the minimum and maximum values across the whole domain.)

\begin{figure}     
    \centering
    \subfigure[]{\includegraphics[width=.48\textwidth]{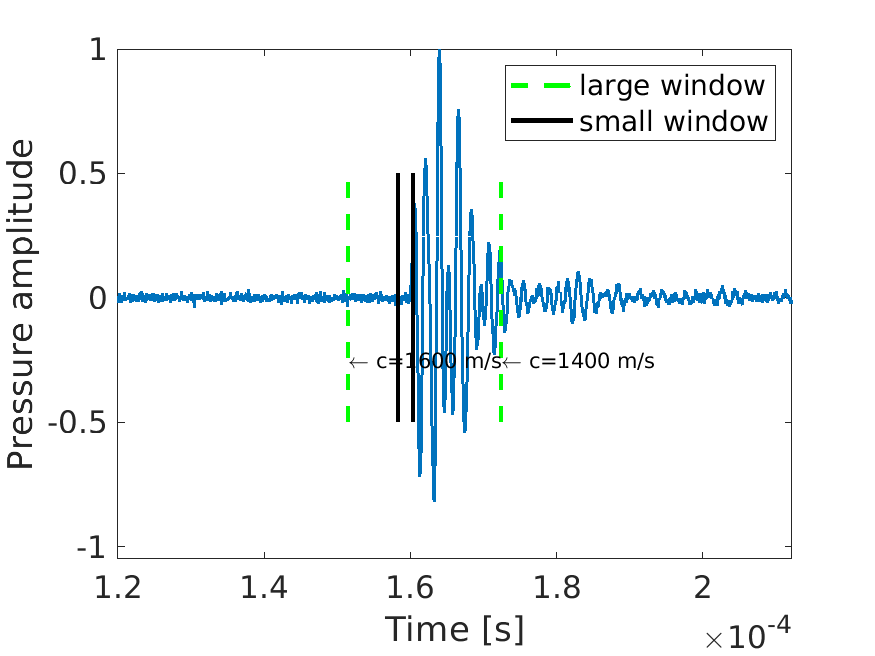}
    \label{fig:time_series_signal}}
    \subfigure[]{\includegraphics[width=.48\textwidth]{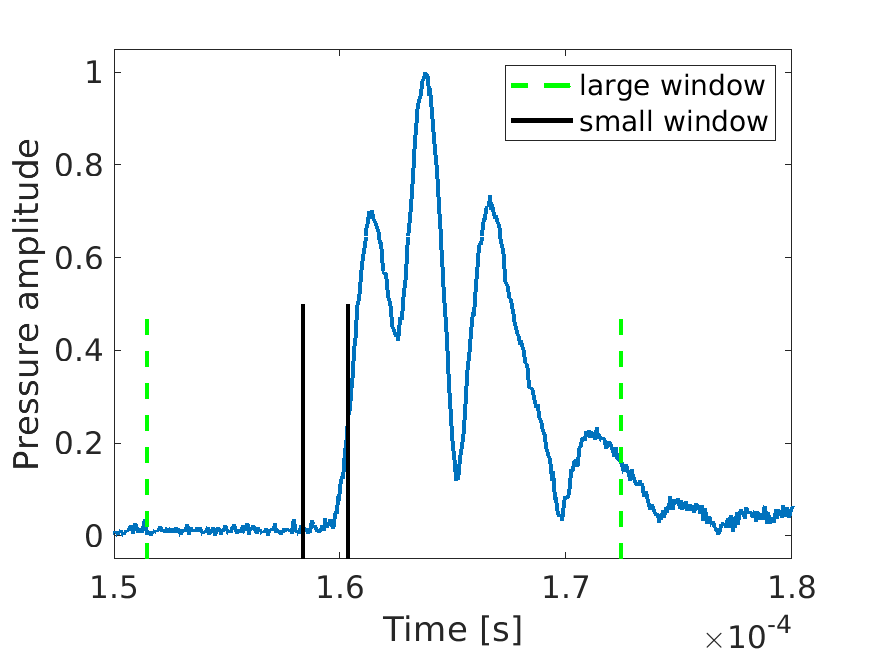}
    \label{fig:signal_envelope}}
    \caption{Large (green) and small (black) time windows. The small window has a size $3 \mu s$, which is almost equal to the temporal duration of the main lobe of the excitation pulse. (a) measured signal (b) the envelope of the measured signal.}
\end{figure}

Figure \ref{fig:time_series_signal} shows a measured signal that is normalised in amplitude. The chosen large window is shown by the green dash lines. $\boldsymbol{y}_{w_l}$ is used to indicate the portion of the signal that inside this large window.

\textbf{Step 3: Small window.} Another window, the \textit{small time window}, that is expected to include the first-arrival of the signal is then chosen. The last point of the small window is chosen as the first time index in the large window for which the amplitude of the envelope of $\boldsymbol{y}_{w_l}$ is greater than a threshold $a_{th} \in \big( 0, 1\big)$. Here, $a_{th}$ is a user-adjusted parameter that is chosen based on our estimation for the noise level in the signal. (Set here to $a_{th}=0.25$.) For choosing the first time index for the small window, we move backward for a time equal to the duration of the main lobe of the excitation pulse. Using the excitation pulse shown in Fig.\ \ref{fig:pulse_time}, the size of the small window was chosen to be $3 \mu$s, and is shown by the black lines in Figs.\ \ref{fig:time_series_signal} and \ref{fig:signal_envelope}. The measured signal inside the small window is denoted $\boldsymbol{y}_{w_s}$, and $i_l$ and $i_r$ denote the indices representing the left and right edges, respectively. Also, the number of time indices in this window is represented by $N_{w_s}$.

\begin{figure}     
    \centering
    \subfigure[]{\includegraphics[width=.48\textwidth]{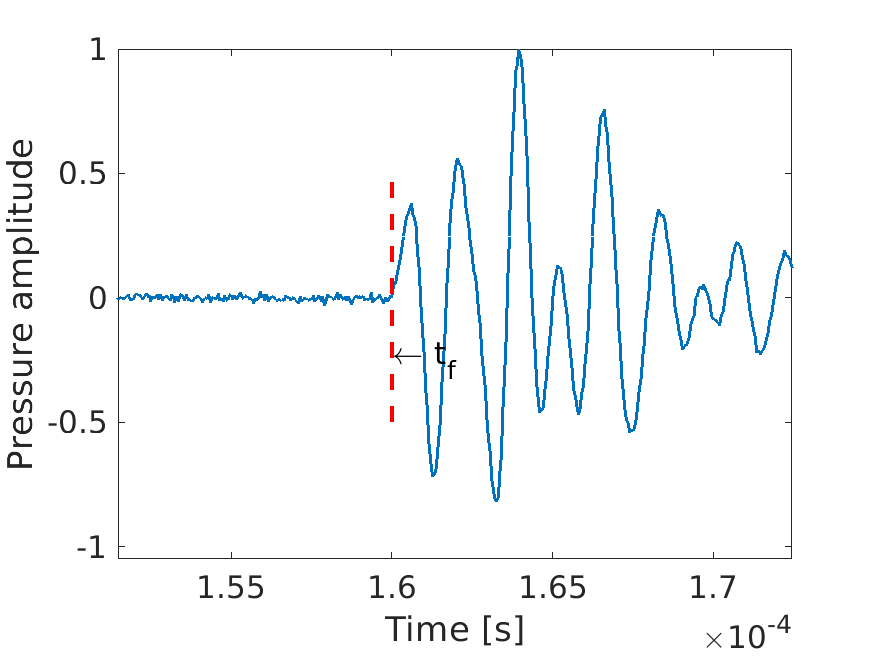}
    \label{fig:TOF_large_window}}
    \subfigure[]{\includegraphics[width=.48\textwidth]{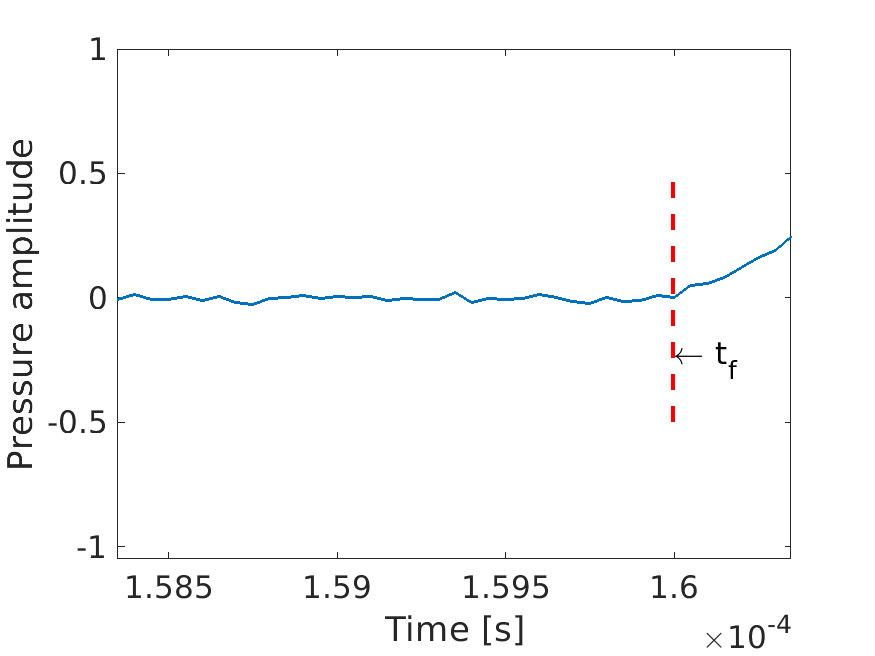}
    \label{fig:TOF_small_window}}
    \caption{The chosen first arrival time $t_f$ from view of (a) the large window and (b) the small window.}
\end{figure}

\textbf{Step 4: Modified Akaike Information Criterion (AIC).} The \textit{Modified AIC} approach is then applied to the portion of the signal that is confined to our small window, $\boldsymbol{y}_{s_w}$, in order to calculate the first arrival of the signal. Using AIC approach, Maeda's formula is used to calculate the AIC of the signal at time index $t_i, \ i \in \left\{i_l,..., i_r \right\}$  \cite{Maeda} 
\begin{align}
AIC(t_i)= i \log \Big( \sigma^2  \big(\boldsymbol{y} (1:t_i)\big)  \Big) + (N_t-i-1)  \log  \Big(\sigma^2 \big(\boldsymbol{y}(t_i+1:N_t) \big)\Big),
\end{align} 
where $\sigma^2$ denotes the variance. Following an approach proposed in \cite{Li3}, we calculate the minimum of AIC in our chosen small time window, and then choose another smaller window containing the minimum, here referred to as the \textit{AIC window}, with time indices $t_{i'}, \ i'\in \left\{  i'_{l},..., i'_{r}  \right\}$ with $1 <= i'_{l} < i'_{r} <= N_{s_w} $. Here, we choose $N_{AIC}$ as the closest integer to $0.25 N_{w_s}$. We then calculate the weights
\begin{align}
\Upsilon (t_{i'})  = \frac{ e^{- \varrho (t_{i'})/2  } } { \sum_{t_{i'}=1}^{N_{AIC}} e^{ - \varrho  (t_{i'})/2 } },
\end{align}
where $ \varrho  (t_{i'})  =  AIC  (t_{i'})   - AIC_{min} $. The first-arrival time $t_f$ is then calculated using
\begin{align}
t_f =  \sum_{t_{i'}=1}^{N_{AIC} }  \Upsilon (t_{i'})  t_{i'}.
\end{align}
Figures \ref{fig:TOF_large_window} and \ref{fig:TOF_small_window} show the calculated first arrival times within the large and small windows respectively. The approach explained above is applied separately to the two sets of data, collected from water alone and with the object present, and the discrepancy between the first arrivals $ \Delta \bar{T}$ is calculated and used as data for solving the UST inverse problem (see Section \ref{subsec:Inverse_problem}).

\begin{thebibliography}{9}


	\bibitem{Hopp1}
	T. Hopp, N. Ruiter, J. C. Bamber, N. Duric and K.W.A. van Dongen (Eds.) 2017 International Workshop on Medical Ultrasound Tomography, Speyer, Germany.
	
	\bibitem{Green}
	J. F. Greenleaf, S. A. Johnson, S. L. Lee, G. T. Herman and E. H. Woo, Algebraic reconstruction of spatial distributions of acoustic absorption within tissue from their two-dimensional acoustic projection, In: Green P.S. (eds) Acoustical Holography, pp. 591-603, 1974.
	
	\bibitem{Op}
	K. J. Opielinski, P. Pruchnicki, P. Szymanowski,
    W. K. Szepieniec, H. Szweda, E. Swis, M. Jozwik,
    M. Tenderenda and M. Bułkowskif, Multimodal ultrasound computer-assisted tomography: An approach to the recognition of breast lesions, COMPUT MED IMAG GRAP vol. 65,pp. 102–114, 2018.
	
	\bibitem{Ruiter}
	N. V. Ruiter, M. Zapf, T. Hopp, R. Dapp, E. Kretzek, M. Birk, B. Kohout, H Gemmeke, 3D ultrasound computer tomography of the breast: A new era?, European Journal of Radiology, vol. 81, Supplement 1, pp. S133-S134, 2012.
	
    \bibitem{Ruiter2}
    H. Gemmeke, T. Hopp, M. Zapf, C. Kaiser, N.V. Ruiter, 3D Ultrasound Computer Tomography: Hardware Setup, Reconstruction Methods and First Clinical Results, NUCL INSTRUM METH A, vol. 873, 2017, pp. 59-65, 2017.

	\bibitem{Duric}
	N. Duric  P. Littrup,  L. Poulo,  A. Babkin, R. Pevzner, E. Holsapple, O. Rama and  C. Glide, Detection of breast cancer with ultrasound tomography: First results with the Computed Ultrasound Risk Evaluation (CURE) prototype, \textit{Med. Phys.} vol. 34, no.2, 2007.

	\bibitem{Li1}
	C. Li, N. Duric, P. Littrup and L. Huang, In-vivo breast sound speed imaging with ultrasound computed tomography, \textit{Ultrasound in Med. \& Biol.}, vol. 35, no. 10, pp. 1615–1628, 2009.
	
	\bibitem{Anderson1} 
	A. H. Anderson and A.C. Kak, Simultaneous algebraic reconstruction techniques (SART): A superior implementation of the ART algorithm, \textit{Ultrasonic Imaging}, vol. 6, no. 1, pp. 81-94, 1984.

	\bibitem{Johnson} 
	S. A. Johnson and J. F. Greenleaf, W. F. Samayoa, F. A. Duck and J. Sjostrand,
    Reconstruction of three-dimensional velocity fields and other parameters by acoustic ray tracing, in \textit{IEEE Proc. Ultrason. Sympos.} pp. 46-51, 1975.
    
    \bibitem{Devaney} 
    A. J. Devaney, Inverse-scattering theory within the Rytov approximation, \textit{Optics Letters}, vol. 6, no. 8, 1981.
   
    \bibitem{Wiskin1}
    J. Wiskin, D. T. Borup, S. A. Johnson and M. Berggren, Non-linear inverse scattering: high resolution quantitative breast tissue tomography, \textit{J. Acoust. Soc. Am.}, vol. 131, no. 5, pp. 3802-13, 2012.

    \bibitem{Wiskin2}
    J. W. Wiskin, D. T. Borup, E. Iuanow, J. Klock, M. W. Lenox, 3-D Nonlinear Acoustic Inverse Scattering: Algorithm and Quantitative Results, \textit{IEEE T ULTRASON FERR} vol. 64, no. 3, 2017.
   
    \bibitem{Beydoun} 
    W. B. Beydoun and A. Tarantola, First Born and Rytov approximations: Modelling and inversion conditions in a canonical example, \textit{J. Acoust. Soc. Am.}, vol. 83, pp. 1045-1055, 1988.
    
    \bibitem{Bellavia}
    S. Bellavia, M. Macconi and S. Pieraccini, Constrained Dogleg methods for nonlinear systems with simple bounds, \textit{Comput. Optim. Appl.}, vol. 53, pp. 771–794, 2012.

    \bibitem{Galanti}
    A. Galanti, The theory of Newton's method, \textit{J   COMPUT APPL MATH}, vol. 124, pp. 25-44, 2000.

    \bibitem{Gu}
    G. Z. Gu, D. H. Li, L. Qi and S. Z. Zhou, Descent directions of quasi-Newton methods for symmetric nonlinear equations. \textit{SIAM J NUMER ANAL}, vol. 40, pp. 1763–1774, 2003.

    \bibitem{Li4}
	D. H. Li and W. Cheng, Recent progress in the global convergence of quasi-Newton methods for nonlinear equations, \textit{Hokkaido Mathematical Journal}, Vol. 36, pp. 729–743, 2007.
	
	\bibitem{Li5}
	D. H. Li and M. Fukushima, A derivative-free line search and global convergence of Broyden-like method for nonlinear equations, \textit{Optimization Meth. \& Soft.}, vol.13, pp.181-201, 2000.

	\bibitem{Cruz}
	W. Cruz, J. Martinez and M. Raydan, Special residual method without gradient information for solving large scale nonlinear systems of equations. \textit{Mathematics of Computation} 75, pp. 1429–1448, 2006.
	
	\bibitem{Li6}
	D.H. Li and M. Fukushima, A globally and superlinear convergent Gauss-Newton based BFGS method for symmetric nonlinear equations. \textit{SIAM J NUMER ANAL} vol. 37, pp. 152–172, 1999.
	
	\bibitem{Griewank}
	A. Griewank, The global convergence of Broyden-like methods with a suitable line search. \textit{J AUST MATH SOC Ser. B} vol. 28, pp. 75–92, 1986.
	
	\bibitem{Powell}
	M. J. D. Powell, A FORTRAN subroutine for solving systems of nonlinear algebraic equations. in “Numerical Methods for Nonlinear Algebraic Equations”, (P. Rabinowitz ed.), Gordon and Breach, London, pp. 115–161, 1970.

	\bibitem{Broyden}
    C. G. Broyden, On the discovery of the “good Broyden” method, \textit{Math. Program., Ser. B} vol. 87, pp. 209–213, 2000.
	
    \bibitem{More}
    J.J. Mor\'e and J. A. Trangenstein, On the global convergence of Broyden’s method, \textit{Mathematics of Computation}, vol. 30, pp. 523–540, 1976.

    \bibitem{Murray}
    W. Murray, Newton-type methods, Stanford University, 2010.

    \bibitem{Zhou}
    W. J. Zhou and D. H. Li, A globally convergent BFGS method for nonlinear monotone equations without any merit functions, \textit{MATHEMATICS OF COMPUTATION}
    Vol. 77, no. 264, pp. 2231–2240, 2008.

    \bibitem{Yuan}
    G. Yuan, Z. Wei, X. Lu, A BFGS trust-region method for nonlinear equations, \textit{Computing}, vol. 92, pp. 317-333, 2011.

    \bibitem{Caselles}
    V. Caselles, R. Kimmel, G. Sapiro, Geodesic active contours. \textit{Int. J. Comput. Vis.}, Vol. 22, no. 1, pp. 61-79, 1997.

    \bibitem{Lin} 
    L. Lin, P. Hu, J. Shi, C. M. Appleton, K. Maslov, L. Li, R. Zhang and L. V. Wang. Single-breath-hold photoacoustic computed tomography of the breast. \textit{Nature communications}, Vol. 9, no.1, 2352, 2018.

    \bibitem{Nyayapathy}
    N. Nyayapathi, J. Xia, Photoacoustic imaging of breast cancer: a mini review of system design and image features, \textit{J. Biomed. Opt.}, vol. 24, no. 12, pp. 121911, 2019.
   
    \bibitem{Runborg}
    O. Runborg, Mathematical models and numerical methods for high frequency waves, \textit{Commun. Comput. Phys.}, vol. 2, pp. 827-880, 2007.

   \bibitem{Anderson2}
    A. H. Anderson and A. C. Kak, Digital ray tracing in two-dimensional refractive fields, \textit{J. Acoust. Soc. Am.}, vol. 72, pp. 1593-1606, 1982.
    
    \bibitem{Anderson3}
    A. H. Anderson, A ray tracing approach to restoration and resolution enhancement in experimental ultrasound tomography, \textit{Ultrasonic Imaging}, vol. 12, pp. 268-291, 1990.
    
    \bibitem{Qu}
    X. Qu, T. Azuma, H. Lin, H. Nakamura, S. Tamano, S. Takagi, S. Umemura, I. Sakuma and Y. Matsumoto, Computational cost reduction by avoiding ray-linking iteration in bent-ray method for sound speed image reconstruction in ultrasound computed tomography, \textit{Jpn. J. Appl. Phys.}, vol. 56, 07JF14, 2017.
    
    \bibitem{Bold}
    G. E. J. Bold T. G. Birdsall,, A top-down philosophy for accurate numerical ray tracing \textit{J. Acoust. Soc. Am.}, vol. 80, pp. 656-660, 1986.

    \bibitem{Ferra}
    V. Farra, Ray tracing in complex media, \textit{Joural of Applied Geophysics}, vol. 30, pp. 55-73, 1993.
    
   	\bibitem{Denis} 
    F. Denis, O. Basset, G. Giminez, Ultrasonic transmission tomography in refracting media: Reduction of refraction artefacts by curved-ray techniques, \textit{IEEE Trans Med Imag}, vol. 14, no. 1, 1995.
   
   \bibitem{Oliveira}
    J. A. R. Oliveira, A. Balatazar, and M. Castel\'an, On ray tracing for sharp changing media, \textit{J. Acoust. Soc. Am.}, vol. 146, no. 3, pp. 1595-1604, 2019.
    
    \bibitem{Cerveny}
    V. \v{C}erven\'y, Ray tracing algorithms in three-dimensional laterally varying layered structures. In: Nolet G. (eds) \textit{Seismic Tomography} in Seismology and Exploration Geophysics, pp. 99-133, Springer, Dordrecht, 1987.
    
    \bibitem{Thurber}
    C. H. Thurber, E. Kissling, Advances in Travel-time calculations for 3D structures, in Advances in seismic event location, pp. 71-99, Kluwer Academic Publishers, 2000.
    
    \bibitem{Norton1}
    S. J. Norton and M. Linzer, Correcting for ray refraction in velocity and attenuation tomography: a perturbation approach, \textit{Ultrasonic Imaging},
    vol. 4, pp. 201-233, 1982. 
    
    \bibitem{Moser}
    T. J. Moser, G. Nolet and R. Snieder, Ray bending revisited, Bulletin of the Seismological Society of America, vol. 82, no. 1, pp. 259-288, 1992.
    
	\bibitem{Huang}
	Q. Huang, Y. Huang, W. Hu, and X. Li, Bezier Interpolation for 3-D Freehand Ultrasound, \textit{IEEE T HUM-MACH SYST}, vol. 45, no. 3, 2015.
	
	\bibitem{Um}
	J. Um and C. Thurber, A fast algorithm for two-point seismic ray tracing, \textit{BSSA}, vol. 77, no.3, pp. 972-986, 1987.
	
	\bibitem{Koketsu}
    K. Koketsu and S. Sekine, Pseudo-bending method for three-dimensional seismic ray tracing in
    a spherical earth with discontinuities, \textit{Geophys. J. Int.}, vol. 132, pp. 339–346, 1998.

    \bibitem{Yang}
    W. Yang, A basic study on two-point seismic ray tracing, December 2003.
		
	\bibitem{Xu}
	T. Xu, F. Li, Z. Wu, C. Wu, E. Gao, B. Zhou, Z. Zhang, G. Xu, A successive three-point perturbation method for fast ray tracing in complex 2D and 3D geological models, \textit{Tectonophysics}, vol. 627, pp. 72–81, 2014.
	
	\bibitem{Wesson}
	R. L. Wesson, Travel-time inversion for laterally inhomogeneous crustal velocity models, \textit{Bull. seism. Soc. Am.}, vol. 61, pp. 729-746, 1971.

	\bibitem{Pereyera}
	V. Pereyera, W. H. L. Lee and H. B. Keller, 
	Solving two-point seismic ray tracing problems in a heterogeneous medium, \textit{BSSA}, vol. 70, no. 1, pp. 79-99, 1980.
	 
	\bibitem{Holm}
    D. Holm, Geometric Mechanics Part 1: Dynamics \& Symmetry, 2nd Edn. 2011, pp. 1-97.
	
	\bibitem{Ali}
    R. Ali, S. Hsieh and J. Dahl, "Open-source Gauss-Newton-based methods for refraction-corrected ultrasound computed tomography," Proc.  SPIE 10955, Medical Imaging 2019: Ultrasonic Imaging and Tomography, 1095508, 15 March 2019.
	
	\bibitem{Klimes}
	L Klime\v{s}, Grid travel-time tracing: second-order method for the first arrivals in smooth media,
	Pure and Applied Geophysics, vol. 148, nos. 3/4, 1996.
	 
	\bibitem{Li2}
	S. Li, M. Jackowski, D. P. Dione, T. Varslot, L. H. Staib, K. Mueller, Refraction corrected transmission ultrasound computed tomography for application in the breast imaging, \textit{Med. Phys.}, vol. 37, no. 5, 2010.
	 
	\bibitem{Tsitsiklis}
	J. N. Tsitsiklis, Efficient algorithms for globally optimal trajectories, \textit{IEEE Trans. Control Syst.}, vol. 40, no. 9, 1995 
	 
	\bibitem{Sethian}
	J. A. Sethian, A fast marching level set method for monotonically advancing fronts, \textit{Proc. Nat. Acad. Sci.}, vol. 93, pp. 1591-1595, 1996.
	
	\bibitem{Lay}
    T. Lay and T. C. Wallace, Modern global seismology, Academic Press, 1995.
   
	\bibitem{Julian}
	B. R. Julian and D. Gubbins, Three-dimensional seismic ray tracing, \textit{J. Geophys.}, vol. 43, pp. 95–113, 1977.
	
	\bibitem{Rawlinson1} 
	N. Rawlinson, G. A. Houseman, and C. D. N. Collins, Inversion of seismic refraction and wide-angle reflection travel-time for three-dimensional layered crustal structure, \textit{Geophys. J. Int.} 145, 381–400, 2001.
	
	\bibitem{Rawlinson2}
	N. Rawlinson, J. Hauser, M. Sambridge, Seismic ray tracing and wavefront tracking in laterally heterogeneous media, Advances in Geophysics, vol. 49, pp. 203-273, 2008.
	
	\bibitem{Sambridge}
	M. S. Sambridge and B. L. N. Kennett, Boundary value ray tracing in a heterogeneous
    medium: A simple and versatile algorithm. Geophys. J. Int. vol. 101, pp. 157–168, 1990.
    
	\bibitem{Huth}
    P. Huthwaite and F. Simonetti, High-resolution imaging without iteration: a fast and robust method for breast ultrasound tomography, J Acoust Soc Am. 130(3):1721-34, 2011. doi: 10.1121/1.3613936.

    \bibitem{Liva}	
    M. P\'erez-Liva, J. L. Herraiz, J. M. Udias, E. Miller, B. T. Cox, and B. E. Treeby, Time domain reconstruction of sound speed and attenuation in ultrasound computed tomography using full wave inversion, J. Acoust. Soc. Am., vol. 141, no. 3, pp. 1595-1604 2017. doi: 10.1121/1.4976688
		
    \bibitem{Bachmann2020}
    E. Bachmann and J. Tromp, Source encoding for viscoacoustic ultrasound computed tomography, J. Acoust. Soc. Am., vol. 147, no. 5, pp. 3221-3235 2020. doi: 10.1121/10.0001191

	\bibitem{Plessix}
	R.-E. Plessix, A review of the adjoint-state method for computing the gradient of a functional with geophysical applications, \textit{Geophys. J. Int.} vol. 167, pp. 495–503, 2006.
	 
	\bibitem{Wang}     
	K. Wang, T. Matthews, F. Anis, C. Li, N. Duric, and M. A. Anastasio, Waveform inversion with source encoding for breast sound speed reconstruction in ultrasound Computed Tomography, \textit{IEEE T ULTRASON FERR}, vol. 62, no. 3, 2015.
	
	\bibitem{Matthews}
	T. P. Matthews, K. Wang, C. Li, N. Duric, and M. A. Anastasio, Regularized Dual Averaging Image Reconstruction for Full-Wave Ultrasound Computed Tomography, \textit{IEEE T ULTRASON FERR}, vol. 64, no. 5, 2017.
	
	\bibitem{Matthews1}
	T. P. Matthews and M. A. Anastasio, Joint reconstruction of the initial pressure and speed of sound distributions from combined photoacoustic and ultrasound tomography measurements,
	\textit{Inverse Problems}, vol. 33, pp. 124002, 2017.

    \bibitem{Lou}
    Y. Lou, W. Zhou, T. P. Matthews, C. M. Appleton and M. A. Anastasio, Generation of anatomically realistic numerical phantoms for photoacoustic and ultrasonic breast imaging, \textit{J Biomed Opt.}, vol. 22, no. 4, pp. 041015, 2017.
    
	\bibitem{Li3}
	C. Li, L. Huangb, N. Duric, H. Zhang, and C. Rowe
	An improved automatic time-of-flight picker for medical
	ultrasound tomography, Ultrasonics., vol. 49, pp. 61-72, 2009.

    \bibitem{Akram}
    J. Akram and D. W. Eaton, Geophysics, vol. 81, no. 2, pp. KS67–KS87, 10.1190/GEO2014-0500.1

   	\bibitem{Maeda}
	N. Maeda, A method for reading and checking phase times in autoprocessing
	system of seismic wave data: Zisin, 1985; 38: 365–379.
 
    \bibitem{Treeby}
    B. E. Treeby and B. T. Cox, k-Wave: MATLAB toolbox for the simulation and reconstruction of photoacoustic wave fields \textit{J. Biomed. Opt.} vol. 15, no. 2, 021314, 2010.
    
    \bibitem{kWave} 
    www.k-wave.org

    \bibitem{Kang}
    H. R. Kang, Computational Color Technology, chapter 9, 2006. doi.org/10.1117/3.660835.ch9.
	
    \bibitem{Pierce1981}
    A.D. Pierce, "Acoustics: An Introduction to its Physical Principles and Applications" \textit{Acoustical Society of America} (1981).

	
   \end{thebibliography}
\end{document}